\documentclass[journal,draftcls,onecolumn,12pt,twoside]{IEEEtran}

\usepackage{amsmath}
\usepackage{amsthm}
\usepackage{amssymb}
\usepackage{graphicx}
\usepackage{subcaption}
\usepackage{url}
\usepackage{epstopdf}
\usepackage{placeins}
\usepackage{epsfig}
\usepackage{bbm}
\usepackage{soul}
\usepackage{amsfonts}
\usepackage{balance}
\usepackage{graphicx}
\usepackage{xfrac}
\usepackage{mathrsfs}
\usepackage[shortlabels]{enumitem}
\usepackage{algorithm} 
\usepackage{algpseudocode}
\usepackage{listings}
\usepackage{footnote}
\usepackage[table]{xcolor}
\usepackage{mathtools,amssymb,nccmath}
\usepackage{cuted}

\makeatother

\DeclareMathOperator*{\argmax}{arg\,max}

\newtheorem{theorem}{Theorem}
\newtheorem{prop}[theorem]{Proposition}
\newtheorem{definition}{Definition}
\newtheorem{lemma}[theorem]{Lemma}
\newtheorem{example}{Example}
\newtheorem{remark}{Remark}
\newtheorem{corollary}[theorem]{Corollary}
\newcommand{\mytilde}{\raise.17ex\hbox{$\scriptstyle\mathtt{‌​\sim}$}}
\usepackage{multirow}

\newcommand\blfootnote[1]{%
	\begingroup
	\renewcommand\thefootnote{}\footnote{#1}%
	\addtocounter{footnote}{-1}%
	\endgroup
}

\begin{document}
	\title{Average Probability of Error for Single Uniprior Index Coding  over Binary-Input Continuous-Output Channels}
	
	\author{%
		\IEEEauthorblockN{Anjana A. Mahesh, Charul Rajput, Bobbadi Rupa, and B. Sundar Rajan\\}
		\IEEEauthorblockA{ Department of Electrical Communication Engineering, Indian Institute of Science, Bengaluru 560012, KA, India \\
			E-mail: \{anjanamahesh,charulrajput,bobbadirupa,bsrajan\}@iisc.ac.in}
	}

	\date{}
	
	{}

	\maketitle
	
	\begin{abstract} 
		Ong and Ho developed optimal linear index codes for single uniprior index coding problems (ICPs) by finding a spanning tree for each strongly connected component of their information-flow graphs, following which Thomas et al. considered the same class of ICPs over Rayleigh fading channels. They developed the min-max probability of error criterion for choosing an index code from the set of bandwidth-optimal linear index codes. Motivated by the above works, this paper deals with single uniprior ICPs over binary-input continuous-output channels. Minimizing the average probability of error is introduced as a criterion for further selection of index codes which is shown to be equivalent to minimizing the total number of transmissions used for decoding the message requests at all the receivers. An algorithm that generates a spanning tree with a lower value of this metric than the optimal star graph is also presented. A couple of lower bounds for the total number of transmissions, used by any optimal index code, are derived, and two classes of ICPs for which these bounds are tight are identified. An improvement of the proposed algorithm for information-flow graphs with bridges and a generalization of the improved algorithm for information-flow graphs obtainable as the union of strongly connected sub-graphs are presented, and some optimality results are derived. 
	\end{abstract}
	
	
	\IEEEpeerreviewmaketitle
	
	\section{INTRODUCTION}
	\label{sec:Intro}
	\blfootnote{A part of the contents of this draft has been published as A. A. Mahesh, C. Rajput, B. Rupa, and B. S. Rajan, ``Average Probability of Error for Single Uniprior Index Coding over Rayleigh Fading Channel,''
	Proceedings of IEEE Information Theory Workshop (ITW 2023), Saint-Malo, France, April 23-28, 2023.}

	A broadcast channel that is very effective for disseminating common content becomes highly inefficient when the users request different content. To address this inefficiency of the broadcast channel over which a server transmits distinct content to a set of receivers, each of which knows a subset of the messages at the server \textit{a priori} as side information, the concept of index coding was introduced by Birk and Kol in \mbox{\cite{BiK}}. In an index coding problem (ICP), a server with access to a set of messages broadcasts over a forward channel to a group of receivers. Each receiver has a subset of the messages at the server available \textit{a priori} as side information and requests,  from the server, another subset of messages, non-intersecting with its side information. The server is informed about the side information of the receivers through a slow backward channel. Index coding aims to satisfy the message requests of all the receivers with the minimum number of server transmissions by utilizing the information of receivers' side information and their data requests. The term ``index coding" is due to Bar-Yossef, Birk, Jayram, and Kol \mbox{\cite{BBJK}}, where the formulation of the ICP involved each receiver demanding a single message. The server knows \textit{a priori} that each receiver is going to demand a single message; only the ``index" of the message demanded is unknown to it.

	A solution of an ICP, which is a set of transmissions from the server that satisfies all the receivers' message requests, is called an index code, and the number of transmissions in an index code is called its length. An index code is said to be linear if all the transmissions are linear combinations of the messages, scalar if the transmissions only involve a single generation of the message, and optimal if the number of transmissions in it is the minimum possible. The original formulation of the ICP  considered a setting where each receiver requests a unique message. This was further explored in \cite{BBJK}, which gave a graphical representation for this problem, called side information graph, and showed that a graph functional called minrank gives the length of an optimal scalar linear index code.

	 Ong and Ho in \cite{OngHo} introduced a class of ICPs, called single uniprior ICPs, where each receiver has a unique message as side information and demands a subset of the messages available at the server. The problem is termed single uniprior since each user knows a ``single" message ``\textit{a priori}". The formulation of the single-uniprior ICP in \mbox{\cite{OngHo, OHL}} was motivated by satellite communications. In this scenario, multiple clients engage in message exchange through a satellite, which functions as a relay. Due to the absence of direct communication links, clients initially transmit their messages to the satellite using an uplink channel. Subsequently, the satellite decodes the messages, re-encodes them, and broadcasts them back to the clients through a downlink channel. In this context, the downlink operation can be likened to a single uniprior ICP, where each client aims to acquire the messages of other clients, armed with knowledge solely about its own single message \textit{a priori}. This class of ICPs was represented graphically using directed graphs called information-flow graphs in \cite{OngHo, OHL}. An algorithm that takes this graph as input and generates its strongly connected components (SCCs) is called the pruning algorithm, and linear index codes for each of these SCCs were also developed based on their spanning trees. Further, it was shown in \mbox{\cite{OngHo, OHL}} that, for this class of ICPs, linear index codes achieve the minimum possible length as opposed to the general case where it was shown by Lubetzky and Stav in \mbox{\cite{LS}} that non-linear index codes outperform linear index codes.

	While the original ICP did not consider transmission errors, noisy index coding was studied in the literature in  \cite{NHV, MR, SR, MCR} among others. A noisy version of the single uniprior ICPs was explored in \cite{TRAR}, where the broadcast channel was assumed to be a Rayleigh fading channel. In this setting, different optimal index codes were found to give different probability of error performances at the receivers, and hence the min-max probability of error criterion to choose a code from amongst optimal index codes was introduced in \cite{TRAR}. This criterion was to find an optimal index code that minimizes the maximum probability of error at the receivers, and an algorithm to determine an optimal index code that satisfies this condition for single uniprior ICPs was also presented in \cite{TRAR}.

	This paper considers single uniprior ICPs where the server broadcasts over binary-input noisy channels with continuous-valued output. Examples of such channels include additive white Gaussian noise channels and fading channels through which binary modulated symbols are assumed to be transmitted. Different receivers are assumed to have similar noise characteristics. For such a setting, we introduce a criterion to choose an index code from the class of optimal index codes minimizing the maximum probability of error at the receivers. In the rest of this paper, the term bandwidth-optimal is used to describe index codes with the minimum possible length. For this setting, we make the following technical contributions. 	
	
	\begin{itemize}
		
		\item Minimizing the average probability of error overall message requests by all the receivers is introduced as a criterion to choose an index code from among the set of bandwidth-optimal index codes satisfying the min-max probability of error criterion. 
		\item We prove that minimizing the total number of transmissions used for decoding the message requests across all the receivers is equivalent to minimizing the average probability of error.  
		
		\item For a given SCC of the information-flow graph representing a single uniprior ICP, we derive a condition for choosing the optimal spanning tree of diameter two. 
		
		\item An algorithm to generate a spanning tree that improves upon the optimal star graph w.r.t. average probability of error while keeping the maximum number of transmissions used for decoding any requested message as two, called Algorithm 1, is presented and its computational complexity is analyzed. 
		
		\item Given the parameters of an information-flow graph, two lower bounds for the total number of transmissions required for decoding the message requests at the receivers are derived. 
		
		\item The tightness of the lower bounds is estimated by comparing their values against the optimal value of the total number of transmissions required for the set of all single uniprior ICPs with five or fewer receivers.
		
		\item Two classes of information-flow graphs, for which Algorithm 1 gives optimal spanning trees w.r.t. the average probability of error, are identified.
		
		\item An improvement of Algorithm 1 for information-flow graphs with bridges and a generalization of the improved Algorithm 1 for information-flow graphs with cut vertices are presented, and some optimality results are derived. 
		
		\item Simulation results supporting the claim that Algorithm 1 gives spanning trees or equivalently index codes with an improved average probability of error performances are also provided. 
		
	\end{itemize}

	The rest of this paper is organized as follows. In section \ref{sec:Sys}, the single uniprior index coding setting considered in this paper as well as relevant existing results, are explained. A criterion to choose an index code from the class of bandwidth-optimal index codes satisfying min-max probability of error criterion in \cite{TRAR} is developed in the following section \ref{sec:AvgPe}. For a chosen SCC of the information-flow graph, how to choose a spanning tree that gives an index code that minimizes the total number of transmissions used in decoding is discussed in section \ref{sec:Algo}. For a given set of parameters of the single uniprior ICP, two lower bounds are derived for the minimum value of the total number of transmissions used in decoding that can be attained by an optimal index code in section \ref{subsec:LB} and two families of information-flow graphs are identified in section \ref{subsec:Opt} for which the lower bounds are tight. This is followed by section \ref{sec:Gen}, which gives a generalization of the results for information-flow graphs, which are obtainable as unions of connected components. Concluding remarks and some directions for future research are given in section \ref{sec:Conc}. Finally, in Appendix A, graph terminologies used in this paper are defined.

	\emph{Notations}: The binary field consisting of the elements $0$ and $1$ is denoted as $\mathbb{F}_2$. For a positive integer $n$, $[n]$ denotes the set $\{1,2,\cdots,n\}$. The set of positive integers is denoted by $\mathbb{Z}^+$. For two sets $A$ and $B$, the notation $A \setminus B$ denotes the set of elements in A that are not in B. A vector $v$ is represented as $\mathbf{v}$. The binomial coefficient represented as $\binom{n}{k}$is equal to $\frac{n!}{k!(n-k)!}$ and $\binom{n}{k}=0$, when $n < 1$ or $n < k$.  For a set $A$, the number of elements in it is represented by $|A|$. The notation $\mathbbm{1}_x$ is used to denote the indicator function which takes the value $1$ when $x$ is true and $0$ otherwise. 
	
	\emph{Graph Notations}: For a graph $\mathcal{G}$, $\mathcal{V}(\mathcal{G})$  denotes the set of vertices in $\mathcal{G}$. The set of arcs(edges) in a directed(undirected) graph $\mathcal{G}$ is denoted as  $\mathcal{E}(\mathcal{G})$. The degree of a vertex $v$ in $\mathcal{G}$ is denoted as $\deg_{\mathcal{G}}(v)$. For a rooted tree $\mathscr{T}$, the root vertex is said to be at level 0, which is denoted as $L_0$. The children of the root vertex are said to be in level 1, denoted $L_1$, their children in level 2, denoted $L_2$, and so on. For a vertex $v \in  \mathcal{V}(\mathcal{G})$, the set of its neighbors in $\mathcal{G}$ is denoted as $N_{\mathcal{G}}(v)$. In a directed graph $\mathcal{G}$, for a pair of vertices $u,v \in  \mathcal{V}(\mathcal{G})$, a double arc is said to exist between $u$ and $v$, if both the arcs $(u,v)$ and $(v,u)$ are present in $\mathcal{E}(\mathcal{G})$ and a single arc is said to exist between them if only one of those arcs is present in  $\mathcal{E}(\mathcal{G})$. For a given graph $\mathcal{G}$, the parameter $\Delta(\mathcal{G})$ is used to denote the maximum degree of a vertex in $\mathcal{V}(\mathcal{G})$, i.e., $\Delta(\mathcal{G}) = \max\limits_{v \in \mathcal{V}(\mathcal{G})}\deg_{\mathcal{G}}(v)$. 
	
	\section{System Model \& Preliminaries}
	\label{sec:Sys}
	
	We consider single uniprior ICPs with the central server accessing a library  $\mathcal{X} = \{x_1,x_2,\cdots, x_n\}$  of $n$ messages, where each message $x_i$ belongs to the binary field $\mathbb{F}_2$, and transmitting to a set of $n$ receivers $\mathcal{R} = \{R_1, R_2,\cdots, R_n\}$. Each of the receivers knows a unique message as side information. Without loss of generality, let us assume that the receiver $R_i$ knows the message $x_i$ \textit{a priori} and demands a subset $\mathcal{W}_i$ of $\mathcal{X} \setminus \{x_i\}$. With the set of receiver demands being denoted as $\mathcal{W} = \{\mathcal{W}_1,\mathcal{W}_2,\cdots,\mathcal{W}_n\}$, this single uniprior ICP is denoted as $\mathcal{I}(n,\mathcal{W})$. 
	For this  ICP $\mathcal{I}(n,\mathcal{W})$, an index code of length $N$ consists of an 
	\begin{enumerate}
		\item an encoding scheme, $\mathcal{F} : \mathbb{F}_2^n \rightarrow \mathbb{F}_2^N$ and
		\item a set of $n$ decoding functions, $\{\mathcal{D}_i\}_{i \in [n]}$, $\mathcal{D}_i : \mathbb{F}_2^N \times \mathbb{F}_2 \rightarrow \mathbb{F}_2^{|\mathcal{W}_i|}$ s.t. at receiver $R_i$, for a realization $\mathbf{x} \in \mathbb{F}_2^n$ of $\mathcal{X}$, $\mathcal{D}_i(\mathcal{F }(\mathbf{x}),x_i) = \mathcal{W}_i$. 
	\end{enumerate}
	
	Further, since linear index codes were shown to be optimal in \cite{OngHo} for single uniprior ICPs, only linear encoding schemes need to be considered, and hence the encoding scheme $\mathcal{F}$ can be represented using an $n \times N$ matrix, $\mathbf{L}$, over $\mathbb{F}_2$. Let the index coded vector be represented as $\mathbf{c} = (c_1,c_2,\cdots,c_N) \in \mathbb{F}_2^N$ such that for a message realization $\mathbf{x} \in \mathbb{F}_2^n$, $\mathbf{c} = \mathbf{x}\mathbf{L}$.   
	
	In this paper, we consider that the server sends each encoded bit separately after binary modulation over a continuous-output channel, and the receiver decodes each of the binary-modulated transmitted symbols separately and then performs index decoding to retrieve the requested message bits. Let the binary modulated symbols corresponding to the index codeword $(c_1,c_2,\cdots,c_N)$ be represented as $(s_1,s_2,\cdots,s_N)$. The channel between the transmitter and each receiver is assumed to be independently and identically distributed, and the channel characteristics are assumed to be known at the receivers. 
	
	%
	For the single uniprior problem $\mathcal{I}(n,\mathcal{W})$, a graphical representation was given in \cite{OngHo} using information-flow graphs, which is defined as follows.

	\begin{definition}
		A single uniprior ICP $\mathcal{I}(n,\mathcal{W})$ is represented using a directed graph called an information-flow graph $\mathcal{G} = (\mathcal{V},\mathcal{E})$, where the vertex set represent the set of receivers, $\mathcal{V} = \{1,2,\cdots,n\}$ and there is an arc from $i$ to $j$ if $R_j$ demands $x_i$, i.e., $\mathcal{E} = \{(i,j) : x_i \in \mathcal{W}_j\}$. 
	\end{definition}

	In \cite{OngHo}, an algorithm called the ``Pruning Algorithm" was presented which took the information-flow graph $\mathcal{G}$ as input and returned its strongly connected components $\mathcal{G}_{\text{sub},i}, \ i \in[N_{\text{sub}}]$ and a collection of arcs $\mathcal{G}^{'}$. For a given strongly connected component (SCC) of an information flow graph on $n$ vertices, representing the messages $x_1,x_2,\cdots,x_n$ as well as the receivers $R_1,R_2,\cdots,R_n$ such that $R_i$ knows $x_i$ \textit{a priori}, the length of an optimal index code was shown to be $n-1$ and an index code of the form $x_1+x_2,\ x_2+x_3,\cdots, x_{i-1}+x_i,\ x_i+x_{i+1}, \cdots, x_{n-1}+x_n$ is sufficient to satisfy the demands of all the $n$ receivers involved in that SCC. Finally an optimal index code for the single uniprior ICP represented by $\mathcal{G}$ was given in \cite{OngHo} using optimal index codes for each SCC,  $\mathcal{G}_{\text{sub},i}$,  $i \in [N_{\text{sub}}]$  and by transmitting the requested message corresponding to each arc in $\mathcal{G}^{'}$. Since \mbox{\cite{OngHo}} considered index coding over noiseless broadcast channels, any index code which minimizes the number of transmissions in it occupies the same bandwidth for binary-modulated transmission and hence, are equivalent to each other in terms of the only metric (bandwidth) under which they can be compared. In the rest of this paper, index codes with the minimum possible length are referred to as `bandwidth-optimal'.

	In \cite{TRAR}, it was shown that when the channel between the source and the receivers is noisy, all bandwidth-optimal index codes are not equivalent in terms of the probability of decoded message error performance at the receivers. To explain this in detail, consider a single uniprior ICP $\mathcal{I}(n,\mathcal{W})$ and a bandwidth-optimal index code of length $N$ for it that gives the index code-word $(c_1, c_2,\cdots,c_N)$ for the message vector $\mathbf{x} \in \mathbb{F}_2^n$. These $N$ coded bits are transmitted as $(s_1,s_2,\cdots,s_N)$ after binary modulation over a continuous-output channel and are estimated independently at each receiver. At a receiver $R_j$, each binary symbol can be assumed to be transmitted over a binary symmetric channel with the probability of error $p_j$, where $p_j$ is determined by the noise characteristics of the channel between the source and receiver $R_j$. Since the channels from the source to different receivers are assumed to be independently and identically distributed, their noise characteristics are identical, and hence it is assumed that $p_1 = p_2 = \cdots = p_n = p$. 
	
	Let the received symbols at $R_j$ be $\{y_i^j\}_{i \in [N]}$. To decode a message, say $x_i \in \mathcal{W}_j$, $R_j$ need not make use of all the $N$ transmissions. Let us assume that $R_j$ uses $l_i^j$ out of the $N$ received symbols $\{y_i^j\}_{i \in [N]}$ to estimate $x_i$. With the estimate of the message bit $x_i$ at $R_j$ being denoted as $\hat{x_i}$, the probability of bit error is given as
	\begin{align*}
		Pr(\hat{x_i} \neq x_i) &= \sum\limits_{k \ odd, \ k  \leq l_i^j}Pr( k \text{ transmissions are in error}) \\
		&=\sum\limits_{k \ odd, \ k \leq l_i^j}\left(\binom{l_i^j}{k}p^k(1-p)^{l_i^j-k}\right)
	\end{align*}
	
	From the above expression, it can be seen that to minimize the probability of error in decoding a particular message, the number of transmissions used in its decoding needs to be minimized. This led to the criterion of choosing a bandwidth-optimal index code that minimizes the maximum number of transmissions used in decoding a requested message at any receiver, which was called the min-max probability of error criterion, in \cite{TRAR}. The paper \cite{TRAR} also gave an algorithm (Algorithm 2 in \cite{TRAR}) to generate bandwidth-optimal index codes satisfying min-max probability of error criterion for every strongly connected component (SCC) of the information-flow graph, $\mathcal{G}$. For an SCC  $\mathcal{G}_{\text{sub},i}$ on $n_i$ vertices, this optimal index code was obtained by coding along the edges of a spanning tree of the complete graph $K_{n_i}$ which minimized the maximum distance between any two vertices connected by an arc in $\mathcal{G}_{\text{sub},i}$. For a chosen spanning tree $\mathscr{T}$, the index code based on it consists of the transmissions $\{x_i +x_j:(i,j) \in \mathcal{E}(\mathscr{T})\}$, i.e., corresponding to every edge in the spanning tree, there is a transmission given by the XOR of the messages representing the end-points of that edge. Since any spanning tree of the complete graph $K_{n_i}$ has $n_i-1$ edges, the optimal index code for an SCC on $n_i$ vertex will have $n_i -1$ transmissions.
	
	For a given ICP $\mathcal{I}(n,\mathcal{W})$ and a chosen index code $\mathcal{C}$, let the maximum number of transmissions used by any receiver to decode a single requested message be denoted as  $l_{max}(\mathcal{C})$. In \cite{TRAR}, it was shown that since there exist spanning trees of diameter two for every complete graph, it is always possible to find an optimal index code $\mathcal{C}$ such that the maximum number of transmissions used to decode a requested message at any receiver, $l_{max}(\mathcal{C})$,  is two.

	\section{Average Probability of Error}
	\label{sec:AvgPe}
	
	For a labeled complete graph $K_m$ on $m$ nodes, Cayley's formula \cite{Cay} gives the number of spanning trees to be $m^{m-2}$. Among these spanning trees, consider the $m$ star graphs with vertex  $i$ as its head for each $i \in [m]$. For the index codes obtained from each of these $m$ star graphs, the maximum number of transmissions required to decode a message bit at any receiver is two. In this section, we look at whether all star graphs give an equal probability of error performance. Then, we show that a star graph is not always optimal in terms of the average probability of error performance and give an algorithm that finds a better tree starting from the star graph. 
	
	For a spanning tree $\mathscr{T}$, an index code $\mathcal{C}_{\mathscr{T}}$ is said to be obtained by coding along the edges of $\mathscr{T}$ if $\mathcal{C}_{\mathscr{T}} =  \{x_i+x_j, \ \forall (i,j) \in \mathcal{E}(\mathscr{T})\}$. Since any spanning tree on $m$ nodes has $m-1$ edges, the index code obtained by coding along its edges is of length $m-1$ and hence is optimal in terms of bandwidth occupied. Further, for all index codes obtained from spanning trees with diameter two, since the maximum number of transmissions used by any receiver is at most two, they are also equivalent with respect to the min-max probability of error criterion. Hence, to further select a code from among the bandwidth-optimal index codes satisfying the min-max probability of error criterion, we need a new metric to evaluate the codes. In this paper, we choose the average probability of error across all message demands at all the receivers as a new criterion to select an index code.
	
	For a single uniprior ICP, $\mathcal{I}(n,\mathcal{W})$ represented by its information-flow graph $\mathcal{G} = (\mathcal{V},\mathcal{E})$ and for a chosen index code $\mathcal{C}$, the average probability of error is defined as
	$$P_{\text{avg},\mathcal{C}} = \frac{1}{E}\sum\limits_{i \in [n]}\sum\limits_{x_j \in \mathcal{W}_i}Pr(x_j^i \neq x_j),$$ where $E = |\mathcal{E}|$, is the total number of arcs in the information-flow graph $\mathcal{G}$ which is same as the total number of demands in the problem. For an estimate $x_j^i$  of the requested message bit $x_j$ at the receiver $R_i$, the probability of error is given by $Pr(x_j^i \neq x_j) = l_j^i \ p(1-p)^{l_j^i-1}$, where $l_j^i$ is the number of transmissions used to estimate $x_j$ at $R_i$. Since we are trying to optimize the performance from amongst the bandwidth-optimal index codes that require at most two transmissions to decode a message at any receiver,  $l_j^i$ takes a value in $\{1,2\}$. Therefore, the average probability of error $P_{\text{avg},\mathcal{C}}$ reduces to 
	
	\begin{align*}
		P_{\text{avg},\mathcal{C}} &= \frac{1}{E}\sum\limits_{i \in [n]}\sum\limits_{x_j \in \mathcal{W}_i}\left(l_j^i \ p(1-p)^{l_j^i-1}\right) \\
		&= \frac{1}{E}\left(\sum\limits_{i \in [n]}\sum\limits_{x_j \in \mathcal{W}_i, \  l_j^i = 1 }p + \sum\limits_{i \in [n]}\sum\limits_{x_j \in \mathcal{W}_i ,\ l_j^i = 2}2p(1-p)\right)	 	 
	\end{align*}
	
	Let the total number of demands in $\mathcal{I}(n,\mathcal{W})$ which require one transmission each to decode, be denoted as $t$, which implies that there are $(E-t)$ demands each of which requires two transmissions. With this, $P_{\text{avg},\mathcal{C}}$ further reduces to 
	
	\begin{align*}
		P_{\text{avg},\mathcal{C}} &= tp + 2(E-t)p(1-p) \\
		&= 2Ep(1-p) - \underbrace{tp\big(2(1-p) - 1\big)}_{\text{Term 2}}
	\end{align*}
	
	For all channels over which information exchange is possible, $p <0.5$, and hence $(1-2p)$ is strictly greater than zero, and hence to reduce the average probability of error, $t$ needs to be increased so as to increase Term 2 above. Since the total number of transmissions used to decode message requests across all receivers is given as $T= 2E-t$, increasing $t$ is equivalent to reducing the total number of transmissions used $T$. Hence, in this paper, we give index codes that minimize the total number of transmissions used from among the class of bandwidth-optimal index codes satisfying the min-max probability of error criterion. Such codes are called optimal index codes, and the spanning trees which result in these optimal index codes are called optimal spanning trees in the remainder of this paper.

	 \section{Choosing an optimal spanning tree}
	 \label{sec:Algo}
	 
	 For a single uniprior ICP represented by the information-flow graph $\mathcal{G}$, since the bandwidth-optimal index code in \cite{OngHo} gives a separate code for each strongly connected component of $\mathcal{G}$, in the rest of this paper, we consider all information-flow graphs to be strongly connected. Further, for an index code $\mathcal{C}_{\mathscr{T}}$ obtained from a spanning tree $\mathscr{T}$, both the notations $l_{max}(\mathcal{C}_{\mathscr{T}})$ as well as $l_{max}(\mathscr{T})$ are used interchangeably to mean the maximum number of transmissions used to decode a single requested message at any receiver while using the index code $\mathcal{C}_{\mathscr{T}}$. 
	 \begin{example}
	 	\label{ex:star}
	 	Consider the single uniprior ICP represented by the information-flow graph, $\mathcal{G}$ shown in Fig. \ref{fig:Ex_star}. For this graph, Algorithm 2 in \cite{TRAR} forms the complete graph  $K_4$ on $4$ labeled nodes and finds a spanning tree of diameter two. There are four possible spanning trees of diameter two for a labeled $K_4$, which are shown in Fig. \ref{fig:Ex1_ST}. The bandwidth-optimal index codes, satisfying the min-max probability of error criterion, obtained from each of these spanning trees are given in Table \ref{Tab:Codes_Ex_star}. The decoding at each of the four receivers and the total number of transmissions used, $T$ for each of these codes, is shown in Table \ref{Tab:Txns_Ex_star}. From Table \ref{Tab:Txns_Ex_star}, it can be seen that even though all four codes are bandwidth-optimal and satisfy the min-max probability of error criterion with the maximum number of transmissions used by any receiver to decode a message in its demand set being two, code $\mathcal{C}_2 $ gives the best average probability of error performance.
	 	
	 	\begin{figure}
	 		\begin{subfigure}{0.25\textwidth}
	 			\centering
	 			\scalebox{0.65}{\includegraphics{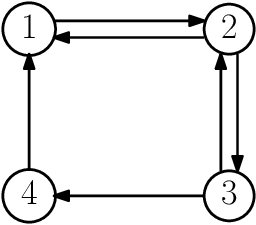}}
	 			\caption{$\mathcal{G}$}
	 			\label{fig:Ex_star}
	 		\end{subfigure}
	 		\hfill
	 		\begin{subfigure}{0.75\textwidth}
	 			\centering
	 			\scalebox{0.65}{\includegraphics{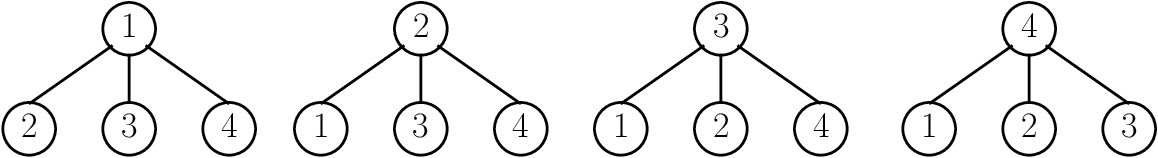}}
	 			\caption{Spanning trees of $K_4$ with diameter two.}
	 			\label{fig:Ex1_ST}
	 		\end{subfigure}
	 		\caption{Information-flow graph and star graphs in Example \ref{ex:star}.}
	 		\label{Example}
	 	\end{figure}
	 	
	 	\begin{table}  	
	 		\centering
	 		\begin{tabular}{|c|c|c|c|c|}
	 			\hline 
	 			& Code $\mathcal{C}_1$ & Code $\mathcal{C}_2$ & Code $\mathcal{C}_3$ & Code $\mathcal{C}_4$ \\ 
	 			\hline 
	 			$c_1$& $x_1+x_2$ & $x_2+x_1$ &$x_3+x_1$  &$x_4+x_1$  \\ 
	 			\hline 
	 			$c_2$& $x_1+x_3$ & $x_2+x_3$ &$x_3+x_2$  &$x_4+x_2$  \\ 
	 			\hline 
	 			$c_3$& $x_1+x_4$ & $x_2+x_4$ &$x_3+x_4$  &$x_4+x_3$  \\ 
	 			\hline 
	 		\end{tabular} 
	 		\caption{Optimal Index Codes from Algorithm 2 in \cite{TRAR} for Example \ref{ex:star}.}
	 		
	 		\label{Tab:Codes_Ex_star}
	 	\end{table}
	 	
	 	\begin{table}
	 		\centering
	 		\begin{tabular}{|c|c|c|c|c|c|}
	 			\hline 
	 			Receiver& Demand & Code $\mathcal{C}_1$ & Code $\mathcal{C}_2$ & Code $\mathcal{C}_3$ & Code $\mathcal{C}_4$ \\ 
	 			\hline 
	 			$R_1$& $x_4$ & $x_1 +c_3$ & $x_1 +c_1+c_3$  & $x_1 +c_1+c_3$ & $x_1 +c_1$ \\ 
	 			\hline 
	 			$R_1$& $x_2$ & $x_1+c_1$ & $x_1 +c_1$ & $x_1 +c_1+c_2$ &  $x_1 +c_1+c_2$\\ 
	 			\hline 
	 			$R_2$& $x_1$ & $x_2+c_1$ & $x_2 +c_1$ & $x_2 +c_1+c_2$ & $x_2 +c_1+c_2$  \\ 
	 			\hline 
	 			$R_2$& $x_3$ & $x_2+c_1+c_2$ & $x_2 +c_2$ & $x_2 +c_2$ & $x_2 +c_2 +c_3$ \\ 
	 			\hline 
	 			$R_3$& $x_2$ & $x_3+c_1+c_2$ & $x_3 +c_2$ & $x_3 +c_2$ & $x_3 +c_2+ c_3$ \\ 
	 			\hline 
	 			$R_4$& $x_3$ & $x_4+c_2+c_3$ & $x_4+c_2+c_3$  & $x_4 +c_3$ & $x_4 +c_3$ \\ 
	 			\hline 
	 			$T$&  & $9$ &  $8$ &  $9$ &  $10$ \\ 
	 			\hline 
	 		\end{tabular} 
	 		\caption{Total Number of transmissions used in decoding for each code in Example \ref{ex:star}.}
	 		\label{Tab:Txns_Ex_star}
	 	\end{table}
	 	\begin{figure} 		
	 		\centering
	 		\scalebox{0.5}{\includegraphics{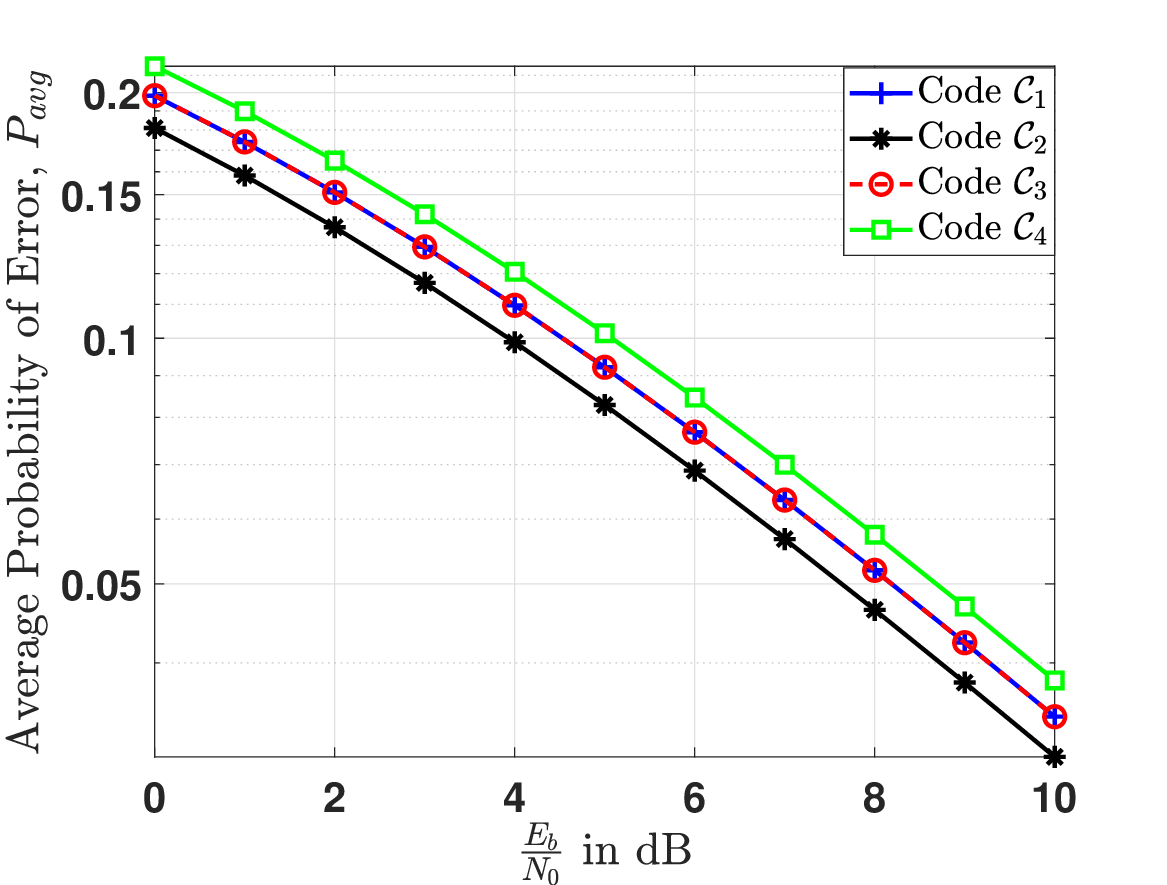}}
	 		\caption{Average probability of error simulation results for Example \ref{ex:star}.}
	 		\label{fig:SimRes}
	 	\end{figure}
	 	
	 	The average probability of error is simulated for each of the four codes $\mathcal{C}_1$, $\mathcal{C}_2$, $\mathcal{C}_3$ and $\mathcal{C}_4$. For this simulation, it is assumed that the broadcast channel is distributed as $\mathcal{CN}(0,1)$ and the additive white noise at each of the receivers is distributed as $\mathcal{CN}(0,N_0)$ and the energy per transmitted bit message is $E_b$, where $\mathcal{CN}(a,b)$ is used to denote complex normal distribution with mean $a$ and variance $b$. The simulation result is given in Fig. \ref{fig:SimRes}, from which it can be seen that the average probability of error performance degrades with increasing value of total number of transmissions used in decoding. 
	 \end{example}

	 \subsection{Optimal Star Graph}
	 For a complete graph $K_m$, the only spanning trees with diameter two are the $m$ star graphs, each with a different vertex as head. Since, from the example above, we saw that not all star graphs perform equally in terms of the average probability of error, we give a criterion for choosing the best star graph. For a strongly connected information-flow graph $\mathcal{G}$ on $n$ vertices,  Algorithm 2 in \cite{TRAR} considers the complete graph $K_{n}$ and returns one of the  $n$ star graphs. A star graph with head vertex $v \in \mathcal{V}(\mathcal{G})$ connected to all the vertices in $\mathcal{V}(\mathcal{G}) \setminus \{v\}$  is denoted as $\mathscr{S}^*_{v}(\mathcal{G})$.

	 \begin{prop}
	 	\label{prop:star}
	 	Consider a single uniprior ICP represented by its information-flow graph $\mathcal{G}$ on $n$ vertices. For an index code based on the star graph $\mathscr{S}^*_{j}(\mathcal{G})$, the total number of transmissions used is $T = 2|\mathcal{E}(\mathcal{G})| - \deg_{\mathcal{G}}(j)$.
	 	
	 	\begin{IEEEproof}
	 		For the single ICP represented by $\mathcal{G}$, the number of arcs in $\mathcal{G}$ given by  $|\mathcal{E}(\mathcal{G})|$ is equal to the number of demands in the ICP. The transmissions in the index code obtained by coding along the edges of a star graph $\mathcal{G}^*_{j}$ are of the form $x_j + x_k$, for all $x_k \in \mathcal{V}(\mathcal{G}) \setminus \{x_j\}$. For this code, the demands that take a single transmission to decode are either when $x_j$ is demanded, represented by outgoing arcs from vertex $j$, or those demanded by the receiver $R_j$ which are represented by incoming arcs to vertex $j$. Hence, the total number of demands taking a single transmission to decode is equal to the sum of in-degree$(j)$ and out-degree$(j)$, which is equal to the degree of $j$ in  $\mathcal{G}$. Every other demand requires two transmissions to decode. 
	 	\end{IEEEproof}
	 	
	 \end{prop}
	 
	 \begin{corollary}
	 	For a single uniprior ICP represented by its information-flow graph $\mathcal{G}$, the star graph which minimizes the average probability of error is the one with vertex $j$ as head, where $j \in \argmax\limits_{ v \in \mathcal{V}(\mathcal{G})}(\deg_{\mathcal{G}}(v))$.
	 \end{corollary}
	 
	 For the information-flow graph in Fig. \ref{fig:Ex1_ST}, vertex $2$ has the maximum degree. Therefore, the star graph with vertex $2$ as the head will use the minimum number of transmissions which can be verified from Table \ref{Tab:Txns_Ex_star} and hence will give the minimum average probability of error.

\begin{figure} 		
	 	\centering
	 	\scalebox{0.65}{\includegraphics{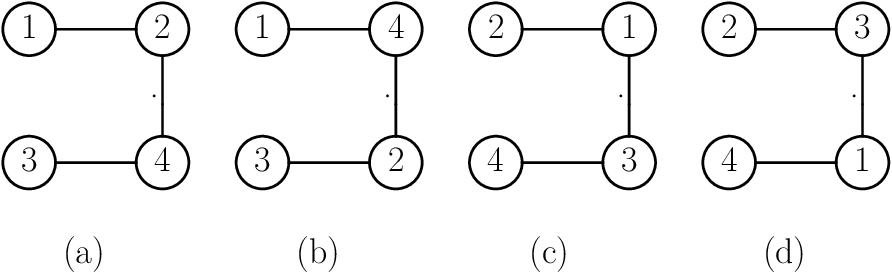}}
	 	\caption{Spanning trees of diameter two other than star graphs for Example \ref{ex:star}.}
	 	\label{fig:Ex1_Trees}
	 \end{figure}

	 While for the complete graph $K_4$, there are only $4$ spanning trees of diameter $2$ as shown in Fig. \ref{fig:Ex1_ST}, for the strongly connected information-flow graph in Fig. \ref{fig:Ex_star}, there are four other trees shown in Fig. \ref{fig:Ex1_Trees} which can be used to generate index codes that satisfy the criterion that the number of transmissions used by any receiver to decode a requested message is at most two. For this example, it can be verified that the index codes obtained by coding along the edges of the trees in Fig. \ref{fig:Ex1_Trees} perform worse than code $\mathcal{C}_2$ given in Table \ref{Tab:Codes_Ex_star} w.r.t. average probability of error. However, this motivates us to look for trees other than star graphs which will minimize the average probability of error further while still using at most two transmissions to decode any requested message. In the following subsection, we give an algorithm to generate a spanning tree that improves upon the star graph in terms of the total number of transmissions used.

	 \subsection{Improving the Optimal Star Graph}
\begin{figure} 	
	 	\centering
	 	\scalebox{0.65}{\includegraphics{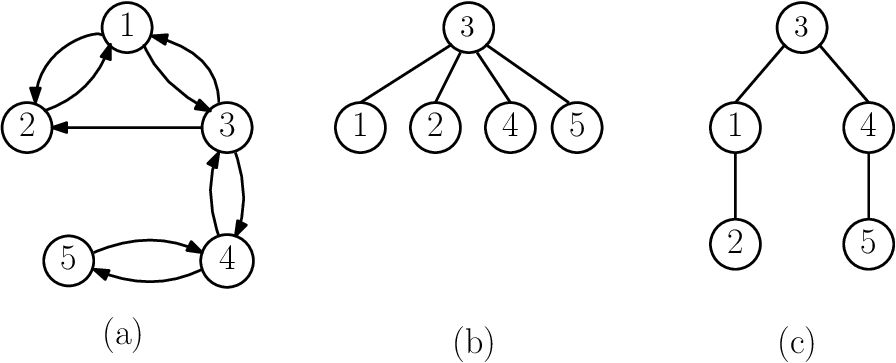}}
	 	\caption{(a) Information-flow graph $\mathcal{G}$, (b) Optimal Star Graph $\mathcal{G}^*_{3}$, and (c) Optimal Spanning Tree $\mathscr{T}$ for Example \ref{ex:Alg1}.}
	 	\label{fig:Ex2_Alg1}
	 \end{figure}

	 \begin{example}
	 	\label{ex:Alg1}
	 	For the information-flow graph, $\mathcal{G}$ in Fig. \ref{fig:Ex2_Alg1}(a), vertex $3$ has the maximum degree, and hence among the $5$ star graphs, the star graph $\mathscr{S}^*_{3}(\mathcal{G})$ with vertex $3$ as head, shown in Fig. \ref{fig:Ex2_Alg1}(b) will give the best average probability of error. The total number of transmissions used in decoding the requested messages when the index code based on the star graph in Fig. \ref{fig:Ex2_Alg1}(b) is transmitted is $T = 13$. Now consider the tree  $\mathscr{T}$ shown in Fig. \ref{fig:Ex2_Alg1}(c). For this tree, the total number of transmissions used is $T=10$. Since the number of arcs in $\mathcal{G}$ is $9$, the minimum value of $T$ is $9$. However, it can be verified that there is no index code of length $N=4$ for which $T=9$. Hence the tree in Fig. \ref{fig:Ex2_Alg1}(c) is optimal w.r.t. average probability of error.
	 \end{example}

	 In the above example, the following two modifications were done on the star graph $\mathscr{S}^*_{3}(\mathcal{G})$ in Fig. \ref{fig:Ex2_Alg1}(b) to obtain the optimal spanning tree in Fig. \ref{fig:Ex2_Alg1}(c).
	 \begin{itemize}
	 	\item The edge $(3,2)$ in $\mathscr{S}^*_{3}(\mathcal{G})$ is removed and the edge $(1,2)$ is added. \label{op1}
	 	\item The edge $(3,5)$ in $\mathscr{S}^*_{3}(\mathcal{G})$ is removed and the edge $(4,5)$ is added. \label{op2}
	 \end{itemize}
	 
	 In $\mathscr{S}^*_{3}(\mathcal{G})$, $(3,2)$ is an edge of the tree, and hence the demand corresponding to $(3,2) \in \mathcal{E}(\mathcal{G})$  takes one transmission to decode in $\mathcal{C}_{\mathscr{S}^*_{3}(\mathcal{G})}$ whereas, $(3,2)$ is no longer an edge of  $\mathscr{T}$ and hence the demand $(3,2) \in \mathcal{E}(\mathcal{G})$ requires two transmissions to decode. But $(1,2)$ is now an edge in the tree, and hence, the demands corresponding to the arcs $(1,2)$ and $(2,1) \in \mathcal{E}(\mathcal{G})$ take one transmission each to decode as opposed to taking two each in $\mathcal{C}_{\mathscr{S}^*_{3}(\mathcal{G})}$. Hence, the modification $\big\{\mathcal{E}(\mathscr{S}^*_{3}(\mathcal{G})) \setminus \{(3,2)\} \big\}\cup \{(1,2)\}$ gives a net reduction of 1 in the total number of transmissions used. Now consider the  operation  $\big\{\mathcal{E}(\mathscr{S}^*_{3}(\mathcal{G})) \setminus \{(3,5)\} \big\}\cup \{(4,5)\}$.  Here, the removal of the edge $(3,5)$ from $\mathscr{S}^*_{3}(\mathcal{G})$ does not affect the total number of transmissions used in decoding as neither $(3,5)$ nor $(5,3)$ is an arc in $\mathcal{G}$ but the addition of the edge $(4,5)$ in $\mathscr{T}$ decreases the number of transmissions needed to decode each of the demands corresponding to $(4,5)$ and $(5,4)$ in $\mathcal{E}(\mathcal{G})$ from two to one. Hence, this operation gives an overall reduction of two to the total number of transmissions used in decoding. 
	 
	 In a directed graph $\mathcal{G} = (\mathcal{V}, \mathcal{E})$, for a pair of vertices $u, \ v \in \mathcal{V}$, the parameter $conn_{\mathcal{G}}(u,v)$ which is used to denote the number of arcs between $u$ and $v$ is defined as 
	 $$conn_{\mathcal{G}}(u,v) = \begin{cases}
	 	1, \text{ if either } (u,v) \in \mathcal{E} \text{ or } (v,u) \in \mathcal{E} \text{ but not both}, \\
	 	2, \text{ if both  }(u,v) \in \mathcal{E} \text{ and } (v,u) \in \mathcal{E}, \\
	 	0, \text{ otherwise. }
	 \end{cases}$$
	 
	 From Example \ref{ex:Alg1}, it can be seen that for a given information-flow graph, $\mathcal{G}$, there is scope for improving a star graph  $\mathscr{S}^*_{j}(\mathcal{G})$ if there exists a vertex $k \in \mathcal{V}(\mathcal{G})$ such that, in $\mathcal{G}$, either
	 \begin{enumerate}
	 	\item $k \notin N_{\mathcal{G}}(j)$ and $N_{\mathcal{G}}(k) = \{l\}$, and/or \label{cond1}
	 	\item $k \in N_{\mathcal{G}}(j)$ with $conn_{\mathcal{G}}(k,j) = 1$ and $N_{\mathcal{G}}(k) \setminus \{j\} = \{l\}$ with $conn_{\mathcal{G}}(k,l)  = 2$. \label{cond2}
	 \end{enumerate} 
	 
	 The improvement is obtained by removing the edge $(j,k)$ from $\mathscr{S}^*_{j}(\mathcal{G})$ and adding the edge $(l,k)$. 
	 As we saw from Proposition \ref{prop:star}, for an index code obtained from star graph $\mathscr{S}^*_{j}(\mathcal{G})$, the number of demands that require one transmission is $\deg_{\mathcal{G}}(j)$. This can be modified to include the edges satisfying conditions \ref{cond1}) and \ref{cond2}) above. For a vertex $j \in \mathcal{V}(\mathcal{G})$, the parameter $adv_{\mathcal{G}}(j)$ is defined as the advantage it gives, i.e., the number of demands which takes one transmission to decode when the transmitted index code is based on a tree obtained by modifying the star graph $\mathscr{S}^*_{j}(\mathcal{G})$ as $\mathscr{S}^*_{j}(\mathcal{G}) \cup \{(l,k)\} \setminus \{(j,k)\} $ for each vertex $k$ satisfying condition \ref{cond1}) or \ref{cond2}) above.
	 
	 \begin{definition}
	 	\label{defn:adv}
	 	Given an information-flow graph $\mathcal{G}$, for a vertex $j \in \mathcal{V}(\mathcal{G})$, its advantage $adv_{\mathcal{G}}(j)$ is defined as $adv_{\mathcal{G}}(j) = \deg_{\mathcal{G}}(j) + p_j + 2o_j$, where, \begin{itemize}
	 		\item $\mathcal{P}_{j}  = \{k \in N_{\mathcal{G}}(j) \text{ s.t. } conn_{\mathcal{G}}(j,k) = 1, \ N_{\mathcal{G}}(k) = \{j,l\} \text{ and } conn_{\mathcal{G}}(k,l) = 2\}$, 
	 		\item $p_j = |\mathcal{P}_{j}|- \frac{1}{2}|\{(k,l) \text{ s.t. } k,l \in \mathcal{P}_j \}$, and
	 		\item $o_j = |\{k  \notin N_{\mathcal{G}}(j) \text{ s.t. } N_{\mathcal{G}}(k) = \{l\} \}|$. 
	 		
	 	\end{itemize}
	 \end{definition}
	 
	 For a demand corresponding to an arc $(i,j) \in \mathcal{E}(\mathcal{G})$, to use one transmission to decode in an index code based on a tree $\mathscr{T}$, it should be an edge in $\mathscr{T}$. Hence the total number of demands which take a single transmission each to decode is the number of arcs $(i,j)$ in $\mathcal{G}$ such that the edge $(i,j)$ is present in $\mathscr{T}$. We define the set $\mathcal{S}_{\mathscr{T}}$ as $\mathcal{S}_{\mathscr{T}} \triangleq \{(i,j) \in \mathcal{E}(\mathcal{G}) \text{ s.t. } (i,j) \in \mathcal{E}(\mathscr{T}) \}$. Hence the number of demands which take one transmission each to decode for the index code based on $\mathscr{T}$ is  $|\mathcal{S}_{\mathscr{T}}|$ which implies that the total number of transmissions used for decoding is $T = 2|\mathcal{E}(\mathcal{G})| -  |\mathcal{S}_{\mathscr{T}}|$. With the notations in place, we propose the following algorithm to perform the modification to the star graph. 

\begin{algorithm}
	\caption*{\textbf{Function 1}: Function to update a tree}
	\begin{algorithmic}[1]
		\Procedure{UpdateTree}{$\mathscr{T}$, $\mathcal{G}$, $\mathcal{V}^{'}$, $i$}
		\While {  $\exists \ j \in \mathcal{V}^{'} $ s.t $N_{\mathcal{G}}(j) \setminus \{i\} = \{k\}$ and $conn_{\mathcal{G}}(j,k) > conn_{\mathcal{G}}(i,j)$} 
		\State $\mathcal{E}(\mathscr{T}) = \big\{\mathcal{E}(\mathscr{T}) \setminus \{(i,j)\} \big\}  \cup \{(k,j)\} $.
		\State $\mathcal{V}^{'} = \mathcal{V}^{'} \setminus \{j,k\}$.
		\EndWhile
		\State \textbf{return} $\mathscr{T}$
		\EndProcedure
	\end{algorithmic}
	\label{func:update}	
\end{algorithm}

\begin{algorithm}
	\caption{Generate a spanning tree which improves upon optimal star graph.}
	
	\begin{algorithmic}[1]
		
	\Require Information-flow graph, $\mathcal{G} = (\mathcal{V},\mathcal{E})$
	\Ensure Tree $\mathscr{T}$
	\State Find the set of all vertices with maximum advantage, $\mathcal{A} := \argmax\limits_{v \in \mathcal{V}}(adv_{\mathcal{G}}(v))$. 
	\State Find the vertices in $\mathcal{A}$ with maximum degree,  $\mathcal{A}_{\Delta} := \argmax\limits_{v \in \mathcal{A}}(\deg_{\mathcal{G}}(v))$. 
	\State Pick an $i$ from $\mathcal{A}_{\Delta}$.
	\State $\mathscr{T} \leftarrow \mathscr{S}^*_{i}(\mathcal{G})$. 
	\If{$\Delta(\mathcal{G}) < adv_{\mathcal{G}}(i)$}
	
	\State $\mathcal{V}^{'}= \mathcal{V} \setminus \{i\}$.
	\State \Call{UpdateTree}{$\mathscr{T}$, $\mathcal{G}$, $\mathcal{V}^{'}$, $i$}
	\EndIf	
	\end{algorithmic}
	\label{alg:ST1}
\end{algorithm}

	\begin{figure} 		
	\centering
	\scalebox{0.65}{\includegraphics{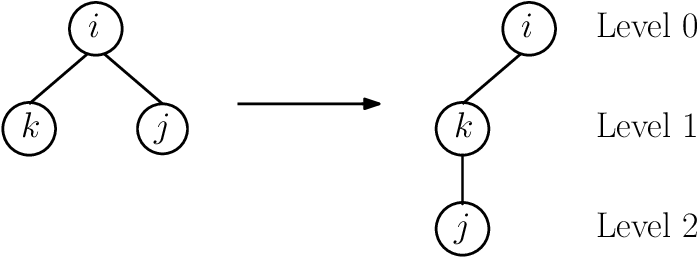}}
	\caption{The proposed modification to star graph in Algorithm \ref{alg:ST1}.}
	\label{fig:Alg1}
\end{figure}

\begin{lemma}
	\label{lem:diam}
	For the index code obtained from the tree $\mathscr{T}$ returned by Algorithm \ref{alg:ST1}, the maximum number of transmissions used by any receiver to decode a single demand is at most 2.
	\begin{IEEEproof}
		For an information-flow graph $\mathcal{G} = (\mathcal{V},\mathcal{E})$, Algorithm \ref{alg:ST1} starts with $\mathscr{T} = \mathscr{S}^*_{i}(\mathcal{G})$, with $i$ being a vertex which gives maximum advantage. Since a star graph has diameter two, for the index code based on $\mathscr{S}^*_{i}(\mathcal{G})$,  $l_{max}(\mathscr{S}^*_{i}(\mathcal{G}))$ is two. It needs to be proved that $l_{max}$ is not increased by the modifications to  $\mathscr{T}$. 
		
		Every modification is of the form $\{\mathcal{E}(\mathscr{T}) \cup \{(k,j)\} \}\setminus \{(i,j)\}$, for a vertex $j \in \mathcal{V} \setminus \{i\}$ which satisfies the condition that $N_{\mathcal{G}}(j) \setminus \{i\} = \{k\}$ and $conn_{\mathcal{G}}(j,k) > conn_{\mathcal{G}}(i,j)$. The operation $\{\mathcal{E}(\mathscr{T}) \cup \{(k,j)\}\} \setminus \{(i,j)\}$ moves the vertex $j$ from Level 1 to Level 2, as shown in Figure \ref{fig:Alg1}, due to which the demands in $\mathcal{G}$ that can require more than two transmissions to decode are of the form $(l,j)$ or  $(j,l) \in \mathcal{E}$ for $l \in \mathcal{V} \setminus \{i,k\}$. Since $N_{\mathcal{G}}(j) \setminus \{i\} = \{k\}$, $N_{\mathcal{G}}(j) \subseteq \{i,k\}$. Hence no such demand exists in $\mathcal{G}$, which can require more than two transmissions to decode.

	\end{IEEEproof}
\end{lemma}

\begin{lemma}
	\label{lem:T}
	For the index code obtained from the tree $\mathscr{T}$ returned by Algorithm \ref{alg:ST1}, the total number of transmissions used in decoding is $T = 2|\mathcal{E}(\mathcal{G})| - adv_{\mathcal{G}}(i)$, where $i$ is a vertex with maximum advantage in $\mathcal{G}$. 
	\begin{IEEEproof}
		From the previous lemma, we know that each demand needs a maximum of two transmissions to decode. As explained earlier, the demand corresponding to an arc $(j, k) \in \mathcal{E}(\mathcal{G})$ needs only one transmission to decode if and only if $(j, k) \in \mathcal{E}(\mathscr{T})$. Therefore, we have $T = 2|\mathcal{E}(\mathcal{G})|- |\mathcal{S}_{\mathscr{T}}|$.  Now we need to prove that $|\mathcal{S}_{\mathscr{T}}|=adv_{\mathcal{G}}(i)$ for the tree $\mathscr{T}$ obtained from Algorithm \ref{alg:ST1}. 
		
		Algorithm 1 starts with a star graph $\mathscr{S}^*_{i}(\mathcal{G})$, for which $|\mathcal{S}_{\mathscr{S}^*_{i}(\mathcal{G})}| = \deg_{\mathcal{G}}(i)$. All the operations performed by Algorithm \ref{alg:ST1} are of the form $\{\mathcal{E}(\mathscr{T}) \cup \{(k,j)\} \}\setminus \{(i,j)\}$,  $\forall j \in V \setminus \{i\}$ which satisfy the condition that $N_{\mathcal{G}}(j)\setminus \{i\} = \{k\}$ and $conn_{\mathcal{G}}(j, k) > conn_{\mathcal{G}}(i, j)$. This condition is equivalent to the following two cases. 
		
		\emph{Case 1}: $j \notin N_{\mathcal{G}}(i)$ and $N_{\mathcal{G}}(j) = \{k\}$ - Since $j$ has only one neighbor $k$ in $\mathcal{G}$, removing the edge $(i,j)$ from $\mathscr{T}$ will not affect the number of transmissions used, whereas, by adding the edge $(k, j)$ to $\mathscr{T}$, $|\mathcal{S}_{\mathscr{T}}|$ will increase by two as both $(j, k)$ and $(k, j) \in \mathcal{E}(\mathcal{G})$ (since $\mathcal{G}$ is strongly connected). The number of vertices satisfying this condition is represented as $o_i$ in Definition \ref{defn:adv}.
		
		\emph{Case 2}: $N_{\mathcal{G}}(j)=\{i,k\},\ conn_{\mathcal{G}}(i,j)= 1 \text{ and } conn_{\mathcal{G}}(j, k) =2$ - Clearly, by removing edge $(i,j)$ from $\mathscr{T}$,  $|\mathcal{S}_{\mathscr{T}}|$ will decrease by one as there is an arc between the vertices $i$ and $j$. However, the addition of the edge $(k,j)$ to $\mathscr{T}$ will increase $|\mathcal{S}_{\mathscr{T}}|$ by $2$ as there are two arcs between $k$ and $j$. Hence, a vertex $j$ satisfying the condition in this case gives a net increment of $1$ in $|\mathcal{S}_{\mathscr{T}}|$. In Definition 2, the number of vertices satisfying this case is represented by $p_i$. 
		
		The while loop in Function \ref{alg:ST1} finds all vertices satisfying either of the two cases above and hence gives $|\mathcal{S}_{\mathscr{T}}|= \deg_{\mathcal{G}}(i) + 2 o_i + p_i$ which is equal to  $adv_{\mathcal{G}}(i)$.
	\end{IEEEproof}
\end{lemma}

\subsection{Complexity Analysis of Algorithm \ref{alg:ST1}}
The step that is most computationally intensive in Algorithm \ref{alg:ST1} is the computation of the parameter $adv_{\mathcal{G}}$ for all the $n$ vertices in the graph $\mathcal{G}$. As described in Definition \ref{defn:adv}, the computation of $adv_{\mathcal{G}}$ for a vertex $v$ involves finding its degree and the values of $p_v$ and $o_v$. Given the adjacency matrix $\mathbf{A}_{\mathcal{G}}$ of the strongly connected graph $\mathcal{G}$, finding the degree of each vertex is of complexity $\mathcal{O}(1)$, and hence finding the degree of all the vertices has a complexity of $\mathcal{O}(n)$. The parameter $o_v$ is the number of vertices outside the neighborhood of the vertex $v$, which has only one neighbor. Assuming that the list of each vertex is stored, this parameter can be calculated simultaneously for all the vertices in $\mathcal{V}(\mathcal{G})$ using the following instructions.
	%
\begin{algorithmic}[1]	
	\For{i = 1 to n}
	\If{$|N_{\mathcal{G}}(i)| == 1$}
	\State Let $k = N_{\mathcal{G}}(i)$
	\State $o_v = o_v + 1, \ \forall v \in \mathcal{V}(\mathcal{G}) \setminus \{i,k\}$
	\EndIf
	\EndFor
\end{algorithmic}
For each vertex $i$ such that $|N_{\mathcal{G}}(i)|= 1$, $n-2$ additions are performed. The loop iterates over all the vertices in $\mathcal{V}(\mathcal{G})$, hence, the computational complexity incurred in calculating $o_v$ for all vertices in $\mathcal{V}(\mathcal{G})$ is $\mathcal{O}(n^2)$. Similarly, the value $p_v$ can be calculated using the following instructions. Here the condition in step \ref{checkA} checks if the $conn_{\mathcal{G}}(i,v)$ is $1$, and the condition in step \ref{checkB} checks if there is only one other neighbor for the vertex $i$, say $k$, and the value of $conn_{\mathcal{G}}(i,k)$ is $2$. Hence, the computation of $p_v$ for all the $n$ vertices in $\mathcal{V}(\mathcal{G})$ has a maximum complexity of $\mathcal{O}(n^2)$. 
\begin{algorithmic}[1]	
	\For{v = 1 to n}
	\State $p_v = 0$
	\State $\mathcal{V}^{'} =  N_{\mathcal{G}}(v)$
	\For{$i \in \mathcal{V}^{'}$}
	\If{$\mathbf{A}_{\mathcal{G}}(i,v) + \mathbf{A}_{\mathcal{G}}(v,i) == 1$} \label{checkA}
	\If{$|N_{\mathcal{G}}(i)| == 2$ \& $\deg_{\mathcal{G}}(i) == 3$} \label{checkB}
	\State $p_{v} = p_v + 1$
	\State $\mathcal{V}^{'} = \mathcal{V}^{'} \setminus \{i \cup N_{\mathcal{G}}(i)\}$
	\EndIf
	\EndIf
	\EndFor
	\EndFor
\end{algorithmic}
Therefore, the overall complexity of computing the parameter $adv_{\mathcal{G}}$ for all the vertices in $\mathcal{V}(\mathcal{G})$, and that of Algorithm \ref{alg:ST1} is $\mathcal{O}(n^2)$

\subsection{Relation to Prim's Algorithm for finding maximum weight spanning trees}
In graph theory, there exist greedy algorithms like Prim's algorithm \cite{Prim} and Kruskal's algorithm \cite{JBK} for finding a minimum-weight spanning tree of a weighted undirected graph, which can be used to find maximum-weight spanning trees as well. For a given information-flow graph $\mathcal{G} = (\mathcal{V}, \mathcal{E})$, we represent the corresponding weighted undirected graph by $\mathcal{G}_{U}^W = (\mathcal{V}_U, \mathcal{E}_U)$, where $\mathcal{V}_U = \mathcal{V}$ and $\mathcal{E}_U = \{(u,v) : u<v, (u,v) \in \mathcal{E} \text{ or } (v,u) \in \mathcal{E}\}$, and a weight function $W:\mathcal{E}_U \rightarrow \mathbb{Z}^+$ defined as $$W((u,v)) = \begin{cases}1, \text{ if either } (u,v) \in \mathcal{E} \text{ or } (v,u) \in \mathcal{E} \text{ but not both}, \\
	2, \text{ if }(u,v) \in \mathcal{E} \text{ and } (v,u) \in \mathcal{E}\end{cases}$$

\begin{figure} 		
	\centering
	\scalebox{0.75}{\includegraphics{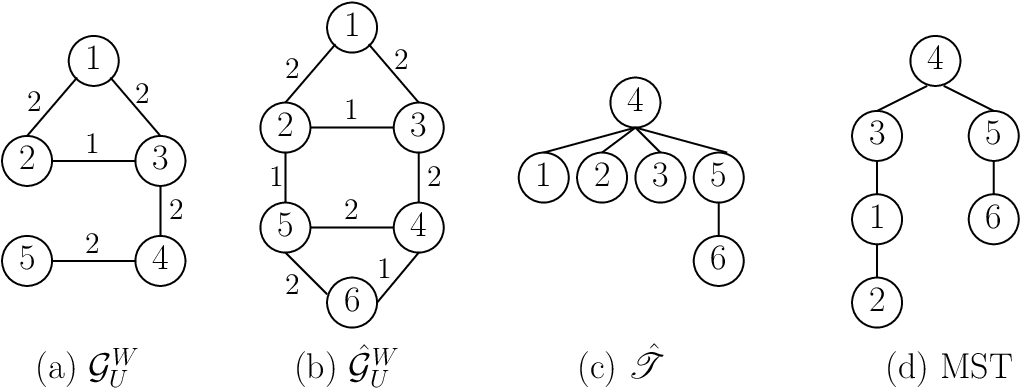}}
	\caption{Illustration of the Insufficiency of Prim's Algorithm.}
	\label{fig:Prims}
\end{figure}

We can run Prim's or Kruskal's algorithm on this weighted undirected graph $\mathcal{G}_U^{W}$ for generating a maximum-weight spanning tree which will ensure the maximum value of $|\mathcal{S}_{\mathscr{T}}|$. However, the index code obtained from the spanning tree thus obtained might satisfy the condition $l_{max} =2$ for some information-flow graphs, whereas for others, it might not. This is illustrated using the following examples. Consider the information-flow graph $\mathcal{G}$ shown in Fig. \ref{fig:Ex2_Alg1}(a). For this directed graph $\mathcal{G}$, the corresponding weighted undirected graph $\mathcal{G}_U^{W}$ is shown in Fig. \ref{fig:Prims}(a). For this graph, running Prim's algorithm starting with any vertex will result in the same spanning tree as the one returned by Algorithm \ref{alg:ST1} shown in Fig. \ref{fig:Ex2_Alg1}(c).

However, if we consider the weighted undirected graph in Fig. \ref{fig:Prims}(b), the optimal spanning tree returned by Algorithm \ref{alg:ST1} is shown in Fig. \ref{fig:Prims}(c) which takes a total of $20$ transmissions for decoding, whereas, running Prim's algorithm on the graph in Fig. \ref{fig:Prims}(b) will result in the spanning tree shown in Fig. \ref{fig:Prims}(d), irrespective of the starting vertex. While the index code based on the tree returned by Prim's algorithm will use only 18 transmissions in total for decoding, it doesn't satisfy the criterion that $l_{max}=2$ as the demand corresponding to the edge $(2,5)$ in $\hat{\mathcal{G}}_U^{W}$ can only be decoded using $4$ transmissions. Hence, in general, for a given information-flow graph, we cannot use Prim's or Kruskal's algorithm for finding the maximum-weight spanning tree of a weighted undirected graph for finding a spanning tree satisfying $l_{max}=2$. 

In the following section, for a given single uniprior ICP, we present a couple of lower bounds for the total number of transmissions used to decode the message requests for any index code and, based on these lower bounds, derive conditions under which the index code based on the tree generated by Algorithm \ref{alg:ST1} is optimal.


\section{Lower Bounds and Optimality Results}
\label{sec:LBOpt}

Let a single uniprior ICP be represented by its information-flow graph $\mathcal{G}$. For this directed graph, let $\mathcal{G}_U$ denote the simplified undirected graph, where simplification involves removing multiple edges between a pair of vertices. Let $\mathcal{D}(\mathcal{G})$ denote the vertex pairs such that there are two arcs between them, i.e., for some ordering on the vertices in $\mathcal{V}(\mathcal{G}), \ \mathcal{D}(\mathcal{G}) = \{(u,v): u < v, \ (u,v), \ (v,u) \in \mathcal{E}(\mathcal{G})\}$ and let $d(\mathcal{G})$ denote the cardinality of the set $\mathcal{D}(\mathcal{G})$, i.e., $d(\mathcal{G}) = |\mathcal{D}(\mathcal{G})|$. 	

\subsection{Lower Bounds}
\label{subsec:LB}

\begin{theorem}
	\label{Thm:LB1}
	For an information-flow graph $\mathcal{G}$ on $n$ vertices, the total number of transmissions used by the receivers to decode their demands is lower bounded as $ T \geq |\mathcal{E}(\mathcal{G})|+|\mathcal{E}(\mathcal{G}_U)|-(n-1)$.
	\begin{IEEEproof}
		The number of demands in the single uniprior problem represented by $\mathcal{G}$ is equal to $|\mathcal{E}(\mathcal{G})|$, each of which takes at least one transmission to decode. For any index code of length $n-1$   which satisfies the condition that any receiver uses at most two transmissions to decode a message request, the number of demands which could take one transmission to decode correspond to at most $n-1$ edges in $\mathcal{G}_U$. Hence, there exists at least $|\mathcal{E}(\mathcal{G}_U)|-(n-1)$ demands, each of which takes one transmission extra over the one already counted in   $|\mathcal{E}(\mathcal{G})|$. Hence, the lower bound. 
	\end{IEEEproof}
\end{theorem}

\begin{theorem}
	\label{Thm:LB2}
	For an information-flow graph $\mathcal{G}$ on $n$ vertices with $d(\mathcal{G})  \geq n-1$, the total number of transmissions used by the receivers to decode their demands is lower bounded as $ T\geq 2(|\mathcal{E}(\mathcal{G})|-n+1)$.
	\begin{IEEEproof}
		From Theorem \ref{Thm:LB1}, we know that the total number of transmissions used by the receivers is at least $|\mathcal{E}(\mathcal{G})|+|\mathcal{E}(\mathcal{G}_U)|-(n-1)$. Since $d(\mathcal{G}) > n-1$, there are at least $d(\mathcal{G}) - (n-1)$ edges, each of which requires an additional transmission to decode. Hence, $T \geq |\mathcal{E}(\mathcal{G})|+|\mathcal{E}(\mathcal{G}_U)|-(n-1) + d(\mathcal{G}) - (n-1)$. Since $|\mathcal{E}(\mathcal{G}_U)|+ d(\mathcal{G}) = |\mathcal{E}(\mathcal{G})|$, we have $ T\geq 2(|\mathcal{E}(\mathcal{G})|-n+1)$.
	\end{IEEEproof}
\end{theorem}

\subsection{Optimality}
\label{subsec:Opt}

\begin{theorem}
	\label{Thm:Opt1}
	For an information-flow graph $\mathcal{G}$ on $n$ vertices, with $d(\mathcal{G})  \leq n-1$, let $i \in \mathcal{V}(\mathcal{G})$ be a vertex with maximum advantage. If $\mathcal{G}$ satisfies the following conditions, then the tree obtained from Algorithm 1 is optimal. 
	\begin{enumerate}
		\item For each $(j,k) \in \mathcal{D}(\mathcal{G})$ such that $j,k \in N_{\mathcal{G}}(i)$, either $N_{\mathcal{G}}(j) = \{i,k\}$ and $conn_{\mathcal{G}}(i,j) = 1$ or  $N_{\mathcal{G}}(k) = \{i,j\}$ and $conn_{\mathcal{G}}(i,k) = 1$.
		\item For each $j \in \mathcal{V}(\mathcal{G}) \setminus  (N_{\mathcal{G}}(i) \cup \{i\})$, $N_{\mathcal{G}}(j) =\{k\}$.
	\end{enumerate}
	\begin{figure} 		
		\centering
		\scalebox{0.65}{\includegraphics{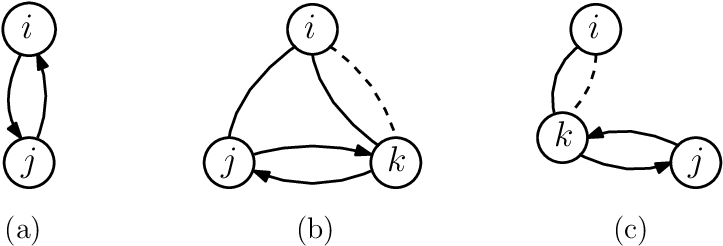}}
		\caption{The possible double-arcs that can exist in $\mathcal{G}$ in Theorem \ref{Thm:Opt1}.}
		\label{fig:Thm1}
	\end{figure}
	\begin{IEEEproof}
		Let $\mathscr{T}$ be the tree obtained from Algorithm \ref{alg:ST1} for the information-flow graph $\mathcal{G}$. Every edge in $\mathscr{T}$ that does not correspond to a double-arc  in $\mathcal{G}$ is of the form $(i,l)$, for some $l \in \mathcal{V}(\mathcal{G}) \setminus \{i\}$. From condition 2) of the theorem, a vertex  $j \notin N_{\mathcal{G}}(i)$ can appear in $\mathcal{G}$ only as shown in Fig. \ref{fig:Thm1}(c) and such a vertex $j$ will be present in Level 2 of the tree $\mathscr{T}$ as shown in Fig. \ref{fig:Alg1}. Hence, for all edges in $\mathscr{T}$ of the form $(i,l)$, $l$ is a neighbor of $i$ in $\mathcal{G}$. 
		
		\emph{Claim}: For every $(j,k) \in \mathcal{D}(\mathcal{G})$, the edge $(j,k)$ is present in $\mathscr{T}$.
		
		The fact that $\forall (i,l) \in \mathscr{T}$, $l \in N_{\mathcal{G}}(i)$, along with the claim, if true, would imply that every edge in the tree $\mathscr{T}$ correspond to an arc in $\mathcal{G}$. Further, since  $d(\mathcal{G})  \leq n-1$, out of the $(n-1)$ edges in the tree $\mathscr{T}$, $d(\mathcal{G})$ edges correspond to double-arcs in $\mathcal{G}$ and the remaining $(n-1-d(\mathcal{G}))$ correspond to single-arcs in $\mathcal{G}$. Thus, the total number of arcs in $\mathcal{G}$ for which a corresponding edge is present in $\mathscr{T}$ is $|\mathcal{S}_{\mathscr{T}}| = 2d(\mathcal{G}) + (n-1-d(\mathcal{G})) = d(\mathcal{G}) + (n-1)$ which implies that the total number of transmissions used in decoding the message requests at the receivers for the index code obtained from $\mathscr{T}$ is $T = 2|\mathcal{E}(\mathcal{G})| - d(\mathcal{G}) - (n-1) = |\mathcal{E}(\mathcal{G})|+|\mathcal{E}(\mathcal{G}_U)|-(n-1)$ which is equal to the lower bound in Theorem \ref{Thm:LB1}. 
		
	\end{IEEEproof}
	
\end{theorem}

\textbf{Proof of Claim}: Satisfying the conditions of the theorem, a double-arc can occur in $\mathcal{G}$ only in the three ways shown in Fig. \ref{fig:Thm1}(a), (b) and (c), where, a solid arc between vertices $u$ and $v$ without direction is used to indicate that either the arc $(u,v)$ or the arc $(v,u)$ exists in $\mathcal{G}$  and a dashed arc indicates that the corresponding arc may or may not be present in $\mathcal{G}$.  
\begin{enumerate}[(a)]
	\item Since Algorithm \ref{alg:ST1} starts with $\mathscr{T}$ as $\mathscr{S}^*_{i}(\mathcal{G})$ and doesn't remove the edge $(i,j)$, this type of double-arc has a corresponding edge in $\mathscr{T}$.
	\item From condition 1) of the theorem, either $N_{\mathcal{G}}(j) = \{i,k\}$ and $conn_{\mathcal{G}}(i,j) = 1$ or $N_{\mathcal{G}}(k) = \{i,j\}$ and $conn_{\mathcal{G}}(i,k) = 1$. If $N_{\mathcal{G}}(j) = \{i,k\}$ and $conn_{\mathcal{G}}(i,j) = 1$, the tree $\mathscr{T}$ will be modified as $\big\{\mathcal{E}(\mathscr{T})  \setminus \{(i, j)\}\big\} \cup \{(k, j)\}$. Similarly if $N_{\mathcal{G}}(k) = \{i,j\}$ and $conn_{\mathcal{G}}(i,k) = 1$, the modification to the tree done by Algorithm \ref{alg:ST1} is $\big\{\mathcal{E}(\mathscr{T})  \setminus \{(i, k)\} \big\} \cup \{(j, k)\}$. In either of these two cases, the edge $(j,k)$ will be added to $\mathscr{T}$, and if an end vertex of the double-arc $(j,k)$ has a double-arc with $i$, it is also represented in $\mathscr{T}$.  
	\item In this case, the vertex $j$ satisfies the condition of the while loop in Function \ref{alg:ST1} and hence the edge $(k,j)$ will be added to the tree $\mathscr{T}$. 
\end{enumerate}

\begin{figure}		
	\centering
	\scalebox{0.65}{\includegraphics{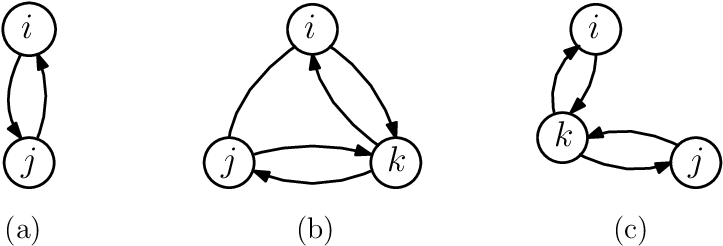}}
	\caption{The possible ways in which a vertex $j \in \mathcal{V}(\mathcal{G}) \setminus \{i\}$ can be present in $\mathcal{G}$ in Theorem \ref{Thm:Opt2}.}
	\label{fig:Thm2}
\end{figure}

\begin{theorem}
	\label{Thm:Opt2}
	For an information-flow graph $\mathcal{G}$ with $d(\mathcal{G})  \geq n-1$, let $i \in \mathcal{V}(\mathcal{G})$ be a vertex with maximum advantage. If all the vertices in $\mathcal{V}(\mathcal{G}) \setminus \{i\}$ satisfy either of the following conditions, then the tree obtained from Algorithm 1 is optimal.
	
	\begin{itemize}
		\item $\forall j \in N_{\mathcal{G}}(i)$, either $conn_{\mathcal{G}}(i,j) = 2$ or $N_{\mathcal{G}}(j) = \{i,k\}$  for a $k \in N_{\mathcal{G}}(i)$ with $conn_{\mathcal{G}}(k,i) = conn_{\mathcal{G}}(j,k) =2$
		\item $\forall j \notin N_{\mathcal{G}}(i)$, $N_{\mathcal{G}}(j) = \{k\}$. 
	\end{itemize}
	\begin{IEEEproof}
		Let the tree obtained from Algorithm \ref{alg:ST1} for the information-flow graph $\mathcal{G}$ be  $\mathscr{T}$. This theorem puts stricter conditions on the type of arcs allowed in the information-flow graph $\mathcal{G}$. A vertex $j \neq i$ can be present in $\mathcal{G}$ only in one of the three ways shown in Fig. \ref{fig:Thm2} and in any of these three cases, the double-arcs will have a corresponding edge in $\mathscr{T}$ and corresponding to the single arc between $i$ and $j$ in Fig. \ref{fig:Thm2}(b), the edge $(i,j)$ will not be present in $\mathscr{T}$. This implies that all tree edges correspond to double-arcs in $\mathcal{G}$ and hence the number of arcs in $\mathcal{G}$ for which a corresponding edge is present in $\mathscr{T}$ is $|\mathcal{S}_{\mathscr{T}}| = 2(n-1)$, which gives the total number of transmissions as $T = 2|\mathcal{E}(\mathcal{G})| - 2(n-1)$ which is equal to the lower bound in Theorem \ref{Thm:LB2}.		
		
	\end{IEEEproof}
\end{theorem}

\begin{example}
	\label{ex:Opt}
	\begin{figure}	
		\centering
		\scalebox{0.65}{\includegraphics{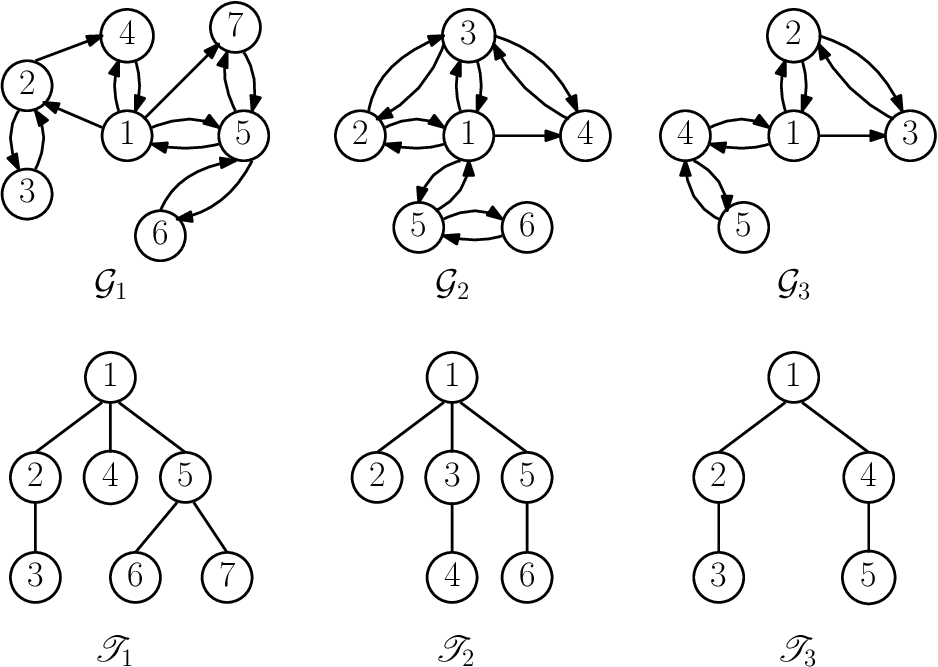}}
		\caption{Information-flow graphs and their corresponding trees obtained from Algorithm \ref{alg:ST1}.}
		\label{fig:Ex_Opt}
	\end{figure}

\begin{table}
	\centering
	\begin{tabular}{|c|c|c|c|}
		\hline 
		Parameter & $\mathcal{G}_1$ & $\mathcal{G}_2$ & $\mathcal{G}_3$ \\ 
		\hline 
		Number of double arcs, $d(\mathcal{G})$ & $5$ & $6$ & $ 4$\\ 
		\hline 
		Number of tree edges, $n-1$ & $6$ & $5$ & $4$ \\ 
		\hline 
		Lower Bound in Theorem \ref{Thm:LB1}  & $15$ & $15$ & $10$  \\ 
		\hline 
		Lower Bound in Theorem \ref{Thm:LB2} & $-$ & $16$ & $10$ \\ 
		\hline 
		Total number of txns used, $T$ & $15$ & $16$ & $10$ \\ 
		\hline
	\end{tabular} 
	\caption{Table showing the values of the lower bounds in Theorems \ref{Thm:LB1} and \ref{Thm:LB2} as well the total number of transmissions used $T$, for the information-flow graphs in Fig. \ref{fig:Ex_Opt}.}
	\label{Tab:Txns_Ex_opt}
\end{table}

Consider the information-flow graphs $\mathcal{G}_1$, $\mathcal{G}_2$, $\mathcal{G}_3$ and their corresponding trees $\mathscr{T}_1$, $\mathscr{T}_2$ and $\mathscr{T}_3$ obtained from Algorithm \ref{alg:ST1} shown in Fig. \ref{fig:Ex_Opt}. Table \ref{Tab:Txns_Ex_opt} gives the value of the lower bounds in Theorems \ref{Thm:LB1} and \ref{Thm:LB2} as well the total number of transmissions used in decoding for the index code based on the tree $\mathscr{T}_i$ for the information-flow graph $\mathcal{G}_i$, for $ i \in [3]$.

\end{example}

\subsection{Tightness of Lower Bounds}
We saw that for information-flow graphs satisfying the conditions in Theorems \ref{Thm:Opt1} and \ref{Thm:Opt2}, the lower bounds in Theorems \ref{Thm:LB1} and \ref{Thm:LB2} are respectively tight. In this subsection, we look at the tightness of the lower bounds by comparing their values against the total number of transmissions used in decoding for an optimal index code for all strongly connected information-flow graphs on five or fewer vertices. With one user in the system, there is no ICP. The server will transmit its demanded message. With two receivers, there is only one single uniprior ICP with the information-flow graph being a double-arc between the vertices $1$ and $2$ for which each of the two receivers will use the server transmission  $x_1 +x_2$ to decode their requested message. This gives the total number of transmissions used as $T=2$, which is equal to both the lower bounds.
\begin{figure}		
	\centering
	\scalebox{0.65}{\includegraphics{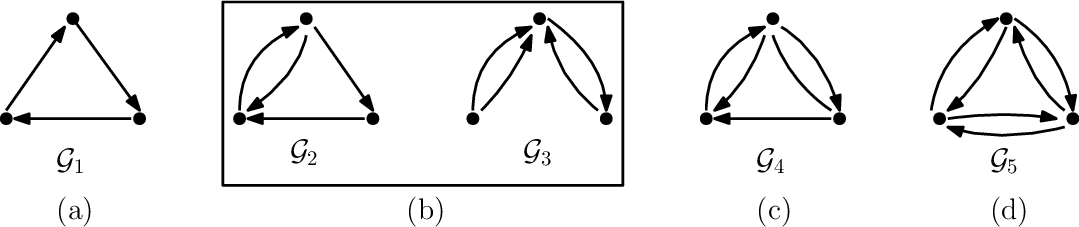}}
	\caption{Strongly connected information-flow graphs on $n=3$ vertices.}
	\label{fig:N=3}
\end{figure}
On $n=3$ unlabeled nodes, there are five strongly connected graphs, shown in Fig. \ref{fig:N=3}, for each of which the star graph with the head as a maximum degree vertex is the optimal tree. The value of the lower bounds and the total number of transmissions used in decoding the index code based on the optimal star graphs are tabulated in Table \ref{Tab:Txns_LB_N=3}.
With $n=4$ unlabeled nodes, there are a total of $83$ strongly connected graphs, and on $n=5$ unlabeled nodes, there are $5048$ strongly connected graphs \cite{OEIS}. Hence, we compare the average values of the total number of transmissions used in decoding an optimal index code and the two lower bounds, where the average is computed over all graphs with a given number of arcs, for which we define the following notations. The average value of the total number of transmissions used in decoding is denoted as $\text{T}_{\text{avg}}$, that of the lower bound in Theorem \ref{Thm:LB1} as $\text{LB1}_{\text{avg}}$ and since the value of the lower bound in Theorem \ref{Thm:LB2} is constant for a given number of arcs, its value is denoted as LB2.

\begin{table}
	\centering
	\begin{tabular}{|c|c|c|c|c|c|}
		\hline 
		Parameter & $\mathcal{G}_1$ & $\mathcal{G}_2$ & $\mathcal{G}_3$ & $\mathcal{G}_4$ & $\mathcal{G}_5$\\ 
		\hline 
		Number of edges, $|\mathcal{E}(\mathcal{G})|$ & $3$ & $4$ & $ 4$ & $5$ & $ 6$\\ 
		\hline 
		Lower Bound in Theorem \ref{Thm:LB1}  & $4$ & $5$ & $4$  & $6$ & $7$\\ 
		\hline 
		Lower Bound in Theorem \ref{Thm:LB2} & $-$ & $-$ & $4$ & $6$ & $8$\\ 
		\hline 
		Total number of txns used, $T$ & $4$ & $5$ & $4$ & $6$ & $8$\\ 
		\hline
	\end{tabular} 
	\caption{Table showing the values of the lower bounds in Theorems \ref{Thm:LB1} and \ref{Thm:LB2} as well the total number of transmissions used $T$, for the information-flow graphs in Fig. \ref{fig:N=3}.}
	\label{Tab:Txns_LB_N=3}
\end{table}
\begin{figure}
	\begin{subfigure}{0.5\textwidth}
		\includegraphics[scale= 0.4]{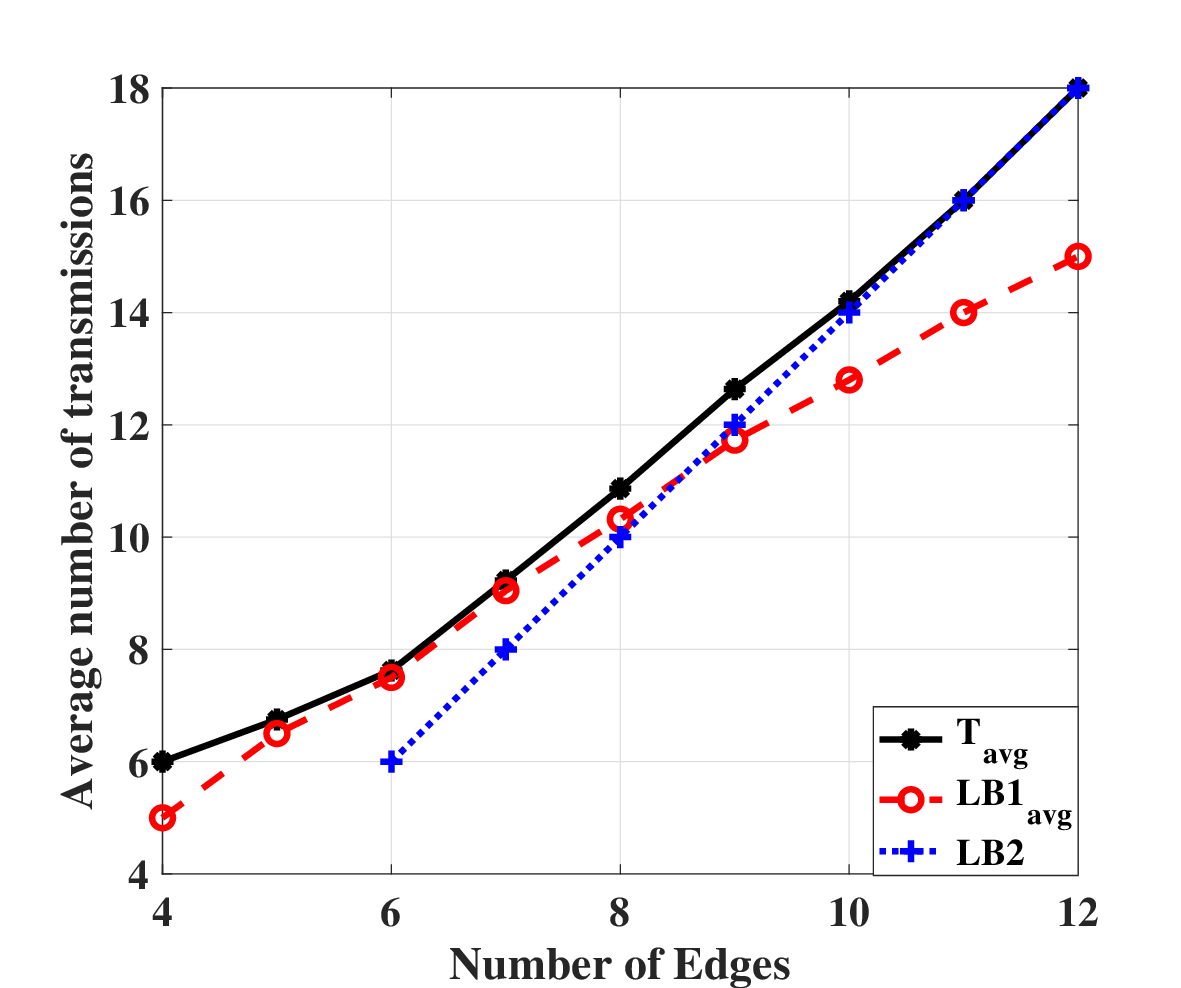}
		\caption{$n=4$ vertices}
		\label{fig:TightLB_n4}
	\end{subfigure}
	\hfill
	\begin{subfigure}{0.5\textwidth}
		\includegraphics[scale= 0.4]{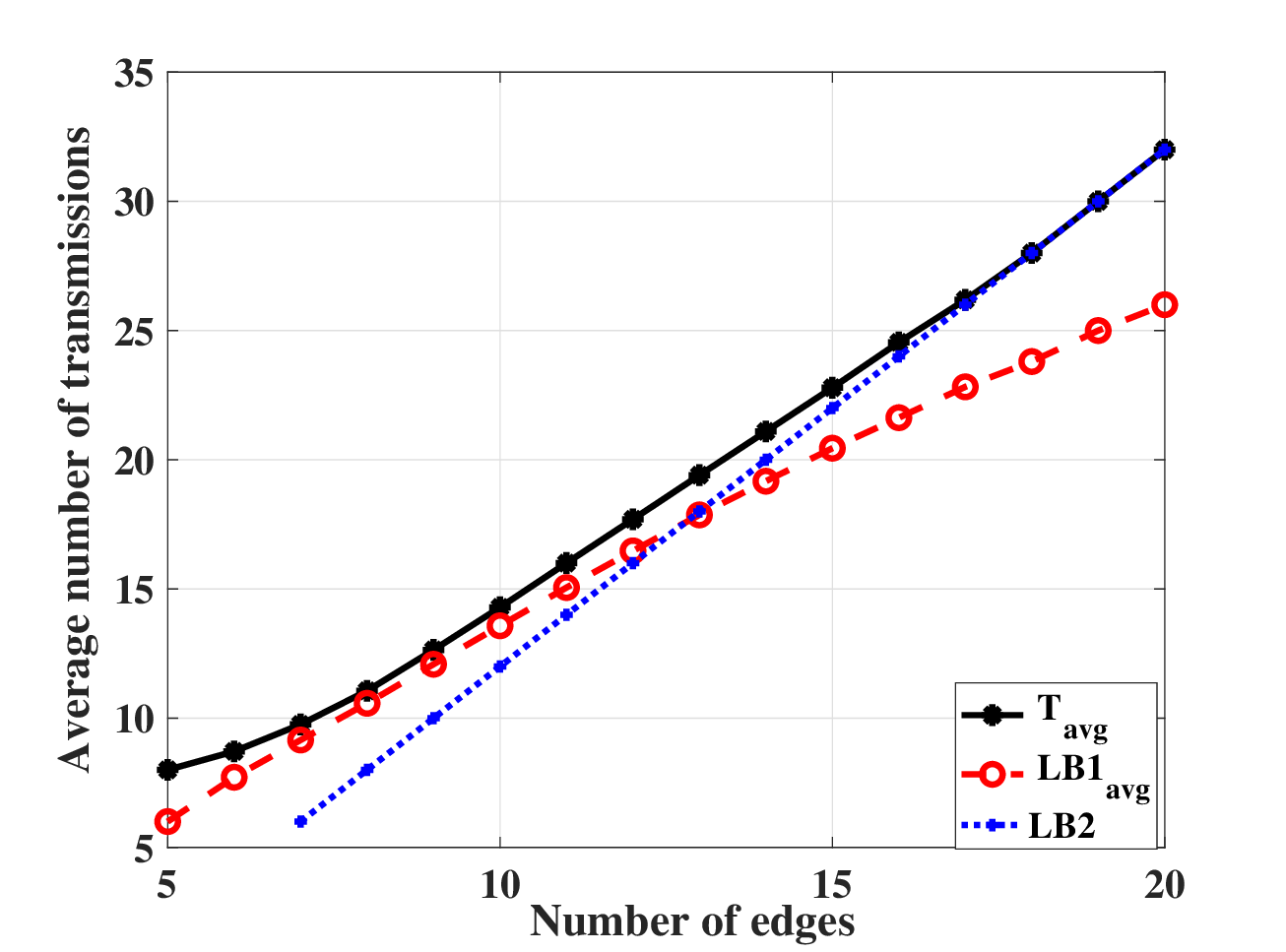}
		\caption{$n=5$ vertices}
		\label{fig:TightLB_n5}
	\end{subfigure}
	\caption{Average value of total number of transmissions used in decoding and two lower bounds for all strongly connected information-flow graphs on $4$ and $5$ vertices.}
	\label{Tightness}
\end{figure}

Plots comparing the values of $\text{T}_{\text{avg}}$, $\text{LB1}_{\text{avg}}$ and LB2 for strongly connected information-flow graphs on $n=4$ and $n=5$ vertices are given in Fig. \ref{fig:TightLB_n4} and Fig. \ref{fig:TightLB_n5}, respectively. Since the lower bound in Theorem \ref{Thm:LB2} is defined only when the number of double-arcs is at least $n-1$, the value of it is plotted only when there exists at least one graph with the given number of arcs satisfying that condition. From these plots, we can observe that when the number of arcs is less, the lower bound in Theorem \ref{Thm:LB1} is closer to the actual value of the total number of transmissions used in decoding, whereas when $|\mathcal{E}(\mathcal{G})|$ is more than $\frac{n(n-1)}{2}$, the value of the lower bound in Theorem \ref{Thm:LB2} gets closer to $\text{T}_{\text{avg}}$. This is to be expected as more the number of arcs, there will be more double-arcs between vertices, and hence there is a higher probability for all the tree edges to correspond to double arcs in the information-flow graph, which is the required condition for the tightness of the lower bound in Theorem \ref{Thm:LB2}. 

\section{Improvements to Algorithm \ref{alg:ST1}}
\label{sec:Imp}	
Consider the following example in which the spanning tree obtained from Algorithm \ref{alg:ST1} can be improved further.

\begin{figure}
	\centering
	\scalebox{0.7}{\includegraphics{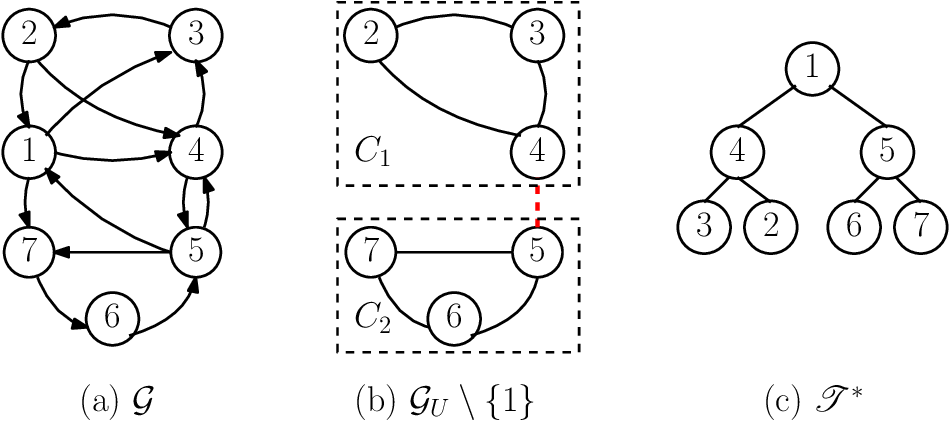}}
	\caption{ Information-flow graph $\mathcal{G}$, $\mathcal{G}_U \setminus \{1\}$ and the optimal tree for $\mathcal{G}$.}
	\label{fig:Ex_alg_ST2}
\end{figure}

\begin{example}
	\label{ex:Alg_ST2}
	Consider the information-flow graph $\mathcal{G}$ in Figure \ref{fig:Ex_alg_ST2}(a). In $\mathcal{G}$, the chosen vertex with maximum advantage is $1$. For this graph, Algorithm \ref{alg:ST1}  cannot improve the star graph $\mathscr{S}^*_{1}(\mathcal{G})$ which takes a total of $T=21$ transmissions to decode the requested messages at all the receivers.  Now, consider the undirected graph $\mathcal{G}_U$ after removing vertex $1$, shown in Figure \ref{fig:Ex_alg_ST2}(b), in which the edge $(4,5)$ is a bridge. Upon removal of the arcs $(4,5)$ and $(5,4)$ from the graph $\mathcal{G} \setminus \{1\}$, it breaks into two components, $C_1$ on vertices $2,3,4$ and $C_2$ on vertices $5,6,7$. The optimal tree $\mathscr{T}$ for the information-flow graph $\mathcal{G}$, shown in Figure \ref{fig:Ex_alg_ST2}(c), was obtained by finding a tree for the components $C_1$ and $C_2$ by running Algorithm $1$  and joining these trees by connecting the bridge vertices to vertex $1$. It can be verified that for the index code based on the tree $\mathscr{T}$, $l_{max}$ is two, and $T = 20$.
	
\end{example}

From the above example, we see that upon removal of a vertex with maximum advantage, if there exists a bridge in the remaining graph, the spanning tree returned by Algorithm \ref{alg:ST1} can possibly be improved further. Using some sample information-flow graphs, we will explain when the new approach will result in a better spanning tree. For a sub-graph $\mathcal{G}_i$ of an information-flow graph $\mathcal{G}$, and a vertex $v \in \mathcal{V}(\mathcal{G})$, the notation $conn_{\mathcal{G}}(v, \mathcal{G}_i)$ is used to denote the number of arcs between $v$ and the vertices in $\mathcal{V}(\mathcal{G}_i)$ in $\mathcal{G}$, i.e., $conn_{\mathcal{G}}(v, \mathcal{G}_i) = \sum\limits_{j \in \mathcal{V}(\mathcal{G}_i)}conn_{\mathcal{G}}(v,j)$.

Let the spanning tree $\mathscr{T}$ returned by Algorithm \ref{alg:ST1}  for an information-flow graph $\mathcal{G}$ be rooted on the vertex $v$. Denote the graph obtained upon removal of the vertex $v$ and the arcs incident on it from $\mathcal{G}$ by $\mathcal{G}^{'}(v)$. Let $\mathcal{V}_B(v)$ denote the bridge vertices (which are endpoints of bridges) in $\mathcal{G}^{'}_U(v)$  and let the arcs in $\mathcal{G}$ corresponding to the bridges in $\mathcal{G}^{'}_U(v)$ be denoted as $\mathcal{B}(v)$, i.e.,  $\mathcal{B}(v) = \{(j,k) \in \mathcal{E}(\mathcal{G}) : (j,k) \text{ is a bridge in }\mathcal{G}^{'}_U(v) \}$. Let the removal of the arcs in $\mathcal{B}(v)$ from  $\mathcal{G}^{'}(v)$ result in $t$ components $C_1,C_2,\cdots, C_t$. A component should have at least two vertices since a single vertex component will be improved while running Algorithm \ref{alg:ST1} for the vertex $v$.  

For a component $C_i$ with a bridge vertex $u_i$, a sub-tree $\mathscr{T}_i$ is generated  by running Algorithm \ref{alg:ST1} on $C_i$ starting with $\mathscr{S}^*_{u_i}(C_i)$. The set of all edges in $\mathscr{T}$ with at least one endpoint being a vertex in $\mathcal{V}(C_i)$ is denoted as $\mathscr{T}_{C_i}$.  To the residual tree $\mathscr{T} \setminus \mathscr{T}_{C_i}$, the sub-tree $\mathscr{T}_i$ is joined by adding an edge $(v,u_i)$.  Generating a sub-tree rooted at a bridge-vertex $u_j$ for a component $C_j$ and attaching the vertex $u_j$ to $v$ will reduce the total number of transmissions used in decoding ($T$)  if and only if $Q(C_j) \coloneqq conn_{\mathcal{G}}(u_j,C_j)-conn_{\mathcal{G}}(v, \left(C_i \setminus \{u_j\}\right) +p_{\mathcal{G}}(u_j) - |\mathcal{P}_{\mathcal{G}}(v) \cap \mathcal{V}(C_i)|$ is greater than zero. The definition for $Q(C_j)$ includes  $p_{\mathcal{G}}(u_j)$ and not $p_{C_j}(u_j)$ because a vertex in $\mathcal{P}_{C_j}(u_j)$  which can give an improvement will also be present in $\mathcal{P}_{\mathcal{G}}(u_j)$. Also, $Q(C_j)$ does not involve a term for  $\mathcal{O}_{C_j}(u_j)$ because the vertices in  $\mathcal{O}_{C_j}(u_j)$ will be present either in $\mathcal{P}_{\mathcal{G}}(v)$ or in $\mathcal{O}_{\mathcal{G}}(v)$ and both these cases will not contribute any additional reduction in $T$. These can be seen from subsequent examples. 

For the information-flow graph $\mathcal{G}$ in Example \ref{ex:Alg_ST2}, the tree returned by Algorithm \ref{alg:ST1} is the star graph $\mathscr{S}^*_{1}(\mathcal{G})$  which has vertex $1$ at level $L_0$ and all other vertices in level $L_1$. For this example, $\mathcal{V}_B(1) = \{4,5\}$ and $\mathcal{B}(1) = \{(4,5),(5,4)\}$. In the final tree $\mathscr{T}^*$, the vertices in the component  $C_2$, except the bridge vertex $5$, have been moved to level $L_2$. The demands corresponding to the arcs between $1$ and the vertices in $N_{\mathcal{G}}(1) \cap (\mathcal{V}(C_2) \setminus \{5\})$ that could be decoded using one transmission each in $\mathscr{S}^*_{1}(\mathcal{G})$ take two transmissions each in $\mathscr{T}^*$. While the component $C_1$ cannot give any reduction in $T$ as  $Q(C_1) = 0$, the  component $C_2$ has $Q(C_2) = 1$ and hence, forming the sub-tree for $C_2$ rooted at vertex $5$ will reduce $T$ by $1$. 

\begin{figure}	
	\centering
	\scalebox{0.7}{\includegraphics{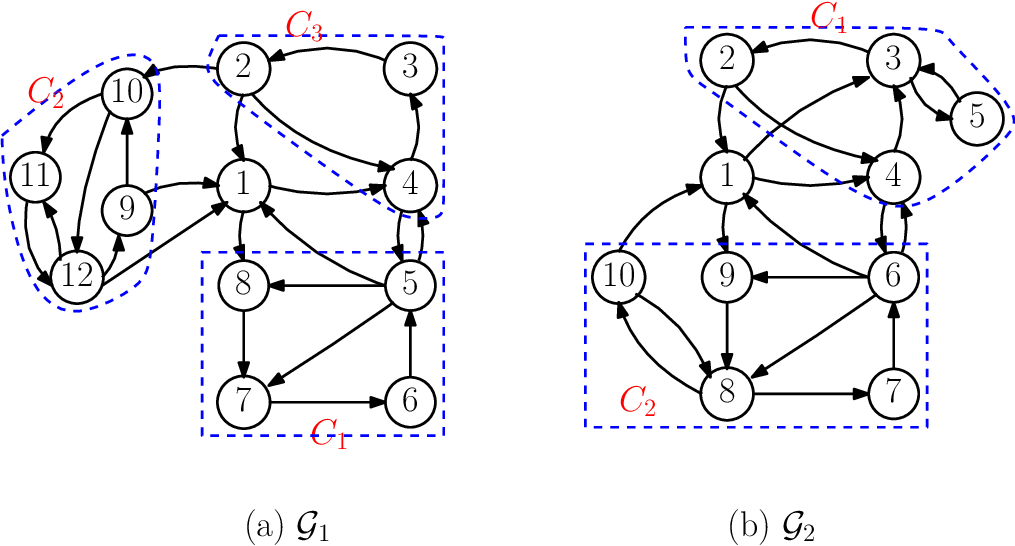}}
	\caption{ Different information-flow graphs for explaining Algorithm A.}
	\label{fig:Ex_alg3_cases}	
	\centering
	\scalebox{0.7}{\includegraphics{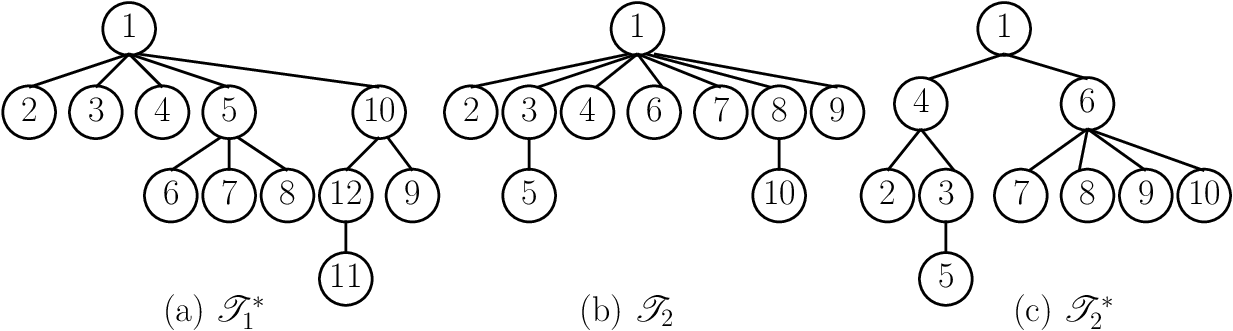}}
	\caption{ Optimal spanning tree for $\mathcal{G}_1$, the spanning tree returned by Algorithm 1 and the optimal spanning tree for $\mathcal{G}_2$.}
	\label{fig:Ex_alg3_cases_trees}
\end{figure}

Now, consider the information-flow graph $\mathcal{G}_1$ given in Figure \ref{fig:Ex_alg3_cases}(a). Upon removal of vertex $1$, which has the maximum advantage, the arcs incident on it and the arcs $(4,5)$, $(5,4)$ and $(2,10)$ corresponding to the bridges, we obtain three components $C_1$, $C_2$ and $C_3$. We have $Q(C_1) = 2$ and hence, forming a sub-graph rooted at vertex $5$ with other vertices in $C_1$ as its children will reduce $T$ by $2$.  Similarly, for the component $C_2$, we have $Q(C_2) = 2$, and hence by forming the sub-tree rooted at vertex $10$, we can reduce $T$ by $2$. Now, consider the component $C_3$, which has two vertices $2$  and $4$ in the bridge vertex set $\mathcal{V}_B$. With either $2$ or $4$, we will get $Q(C_3) = 1$. However, if we create a sub-tree for $C_3$ rooted at vertex 4, in the final tree $\mathscr{T}$, vertex $2$ will be at a distance $3$ from the vertex $10$, which implies that the demand corresponding to the arc $(2,10)$ in $\mathcal{G}_1$ will take three transmissions to decode. Hence, the index code based on this spanning tree will not satisfy $l_{max} = 2$. 

Similarly, if we consider the sub-tree for $C_3$ rooted at vertex $2$, then vertex $4$ will be at a distance $3$ from vertex $5$, and both the demands $(4,5)$ and $(5,4)$ will take three transmissions each for decoding. Hence, the number of transmissions used in decoding the demands in a component $C_i$ can be reduced by forming a sub-tree while maintaining $l_{max} =2$ only if $\mathcal{V}(C_i) \cap \mathcal{V}_B$ is a singleton set.  While the spanning tree $\mathscr{S}^*_{1}(\mathcal{G}_1)$ returned by Algorithm 1 for $\mathcal{G}_1$ has $T= 40$, the optimal spanning tree, $\mathscr{T}_1^*$, given in Figure \ref{fig:Ex_alg3_cases_trees}(a) has $T =36$.

Consider $\mathcal{G}_2$ shown in Figure \ref{fig:Ex_alg3_cases}(b). The spanning tree, $\mathscr{T}_2$, returned by Algorithm \ref{alg:ST1} for this information-flow graph is shown in Figure \ref{fig:Ex_alg3_cases_trees}(b). The vertex with maximum advantage in $\mathcal{G}_2$ is $1$ and has $\mathcal{P}_{1} = \{10\}$. Upon removal of  the vertex $\{1\}$, the arcs incident on it and the arcs  $(4,6)$ and $(6,4)$ corresponding to the bridge, we get  two components $C_1$ and $C_2$. In the tree $\mathscr{T}_1$, we see that vertex $5$ is in level $L_2$ connected to vertex $3$ and it contributes to $\mathcal{O}_{\mathcal{G}_2}(1)$ in $adv_{\mathcal{G}_2}(1)$. Upon generation of the sub-tree for $C_1$ and attaching the bridge vertex $4$ to  $1$, the vertex $5$ is moved to $L_3$. However, this does not increase $l_{max}$. The vertex $10 \in \mathcal{P}_1$ is also in level $L_2$ in $\mathscr{T}_2$. In $C_2$, it contributes to $\mathcal{O}_{C_2}(6)$ and hence is moved to $L_3$ as a child of vertex $8$. This increases $l_{max}$ to $3$. Hence, after forming the star graph $\mathscr{S}^*_{6}(C_2)$, we have to ensure that all the vertices in $C_2$ which are in $N_{\mathcal{G}_2}(1)$ cannot be moved further  down which implies that we have to set $\mathcal{V}^{'}$ in  \Call{UpdateTree}{} as $\mathcal{V}(C_2) \setminus (N_{\mathcal{G}_2}(1) \cup \{6\})$. The optimal spanning tree $\mathscr{T}_2^*$ for $\mathcal{G}_2$ is given in Figure \ref{fig:Ex_alg3_cases_trees}(c). 

In the information-flow graph $\mathcal{G}$ in Figure \ref{fig:Ex_alg_ST2}(a), there are two vertices $1$ and $4$ with maximum advantage. However, if we pick vertex $4$, there are no bridges in  $\mathcal{G}^{'}_U (4)$ and hence, no improvement can be obtained over the star graph $\mathscr{S}^*_{4}(\mathcal{G})$ returned by Algorithm \ref{alg:ST1}. Therefore, the parameter $adv_{\mathcal{G}}$, which is maximized to select the root vertex in Algorithm \ref{alg:ST1}, should be updated to include the advantage obtained from each of the components. To reflect this, we define the modified advantage, $\widehat{Adv}_{\mathcal{G}}$. 
\begin{definition}
	\label{defn:mod_adv}
	For an information-flow graph $\mathcal{G}$ and a vertex $v \in \mathcal{V}(\mathcal{G})$, let the graph $\mathcal{G}^{'}(v) \setminus \mathcal{B}(v)$ break into $m \geq 2$ components, out of which let $t$ components $C_1,C_2,\cdots,C_t$ be such that $\mathcal{V}(C_j) \cap \mathcal{V}_B(v)$ is a singleton set $\{u_j\}$, $j \in [t]$. The modified advantage obtained by $v$ is $\widehat{Adv}_{\mathcal{G}}(v) = adv_{\mathcal{G}}(v) +  \sum_{i \in [t]}Q(C_i){\mathbbm{1}_{Q(C_i) > 0}}$, where 
	$Q(C_i)$=$conn_{\mathcal{G}}(u_i,C_i)-conn_{\mathcal{G}}(v, \left(C_i \setminus \{u_i\}\right) +p_{\mathcal{G}}(u_i) - |\mathcal{P}_{\mathcal{G}}(v) \cap \mathcal{V}(C_i)|$.
\end{definition}
Clearly, the  information-flow graphs for which further improvement of the spanning tree returned by Algorithm 1  is possible are exactly those that satisfy the following condition: \begin{equation}
	\label {condA}
	\max\limits_{v \in \mathcal{V}(\mathcal{G})}\widehat{Adv}_{\mathcal{G}}(v) > \max\limits_{v \in \mathcal{V}(\mathcal{G})}{adv}_{\mathcal{G}}(v).
\end{equation} A modified version of Algorithm 1 which incorporates the modified advantage $\widehat{Adv}$, is described below. 

\begin{algorithm}
	\caption{Modified version of \textbf{Algorithm \ref{alg:ST1}} which incorporates $\widehat{Adv}$}
	
	\begin{algorithmic}[1]
		
		\Require Information-flow graph, $\mathcal{G} = (\mathcal{V},\mathcal{E})$
		\Ensure Tree $\mathscr{T}$
		\State Find the set of all vertices with maximum modified advantage, $\mathcal{A} \coloneqq \arg\max\limits_{v \in \mathcal{V}}(\widehat{Adv}_{\mathcal{G}}(v)) $. 
		\State Find the vertices in $\mathcal{A}$ with maximum degree,  $\mathcal{A}_{\Delta} \coloneqq \arg\max\limits_{v \in \mathcal{A}}(\deg_{\mathcal{G}}(v))$. 
		\State Pick an $i$ from $\mathcal{A}_{\Delta}$.
		\State $\mathscr{T} \leftarrow \mathscr{S}^*_{i}(\mathcal{G})$. 
		\If{$\Delta(\mathcal{G}) < \widehat{Adv}_{\mathcal{G}}(i) $}
		
		\State $\mathcal{V}^{'}= \mathcal{V} \setminus \{i\}$.
		\State \Call{UpdateTree}{$\mathscr{T}, \mathcal{G}, \mathcal{V}^{'},i$}
		\EndIf	
	\end{algorithmic}
	\label{alg:ST2}
\end{algorithm}

\subsection{Proposed Algorithm and Results}
\label{sec:MainResults}
For information-flow graphs satisfying the condition in \eqref{condA}, we propose Algorithm \ref{alg:ST3}, which outputs a spanning tree that will result in an index code with a strictly lower value of the total number of transmissions used in decoding, compared to the index code based on the spanning tree returned by Algorithm \ref{alg:ST1}. 
\begin{algorithm}
	\caption{Algorithm  to generate an improved spanning tree for information-flow graphs  satisfying the condition \eqref{condA} }
	\label{alg:ST3}
	\begin{algorithmic}[1]
		
		\Require Information-flow graph, $\mathcal{G} = (\mathcal{V},\mathcal{E})$
		\Ensure Tree $\mathscr{T}$
		\State Run \textbf{Algorithm \ref{alg:ST2}} on $\mathcal{G}$ to get the tree $\mathscr{T}$.
		\State Let $v \in \mathcal{V}$ be the root vertex of $\mathscr{T}$. 
		\State Let $\mathcal{B}(v) = \{(j,k) \in \mathcal{E}(\mathcal{G}) : (j,k) \text{ is a bridge in }\mathcal{G}^{'}_U(v) \}$.
		\State Let there be $t$ components $C_1,C_2,\cdots,C_t$, in $\mathcal{G}^{'}(v) \setminus \mathcal{B}(v)$, such that $\mathcal{V}(C_j) \cap \mathcal{V}_B(v)$ is a singleton set $\{u_j\}$, $j \in [t]$. 
		\For{ j = 1 : t}
		\If{$Q(C_j) > 0$}
		\State $\mathscr{T} _j=  \mathscr{S}^*_{u_j}(C_j)$
		\State $\mathcal{V}^{'} = \mathcal{V}(C_j) \setminus (N_{\mathcal{G}}(v) \cup \{u_j\})$
		\State $\mathscr{T} _j$ = \Call{UpdateTree}{$\mathscr{T} _j,\mathcal{G},\mathcal{V}^{'},u_j$}
		\State $\mathscr{T}= \left(\mathscr{T} \setminus \mathscr{T}_{C_j}\right) \cup \{(v,u_j)\} \cup \mathscr{T} _j$
		\EndIf
		\EndFor	
		
	\end{algorithmic}
\end{algorithm}

\begin{remark}
	Algorithm \ref{alg:ST3} can be run recursively inside each component if it satisfies the condition in \eqref{condA}	to further reduce the total number of transmissions used in decoding.
\end{remark}

\begin{lemma}
	For an information-flow graph $\mathcal{G}$, for the index code obtained from the spanning tree $\mathscr{T}$ rooted at vertex $v$ returned by Algorithm \ref{alg:ST3}, the total number of transmissions used in decoding the message requests is $T = 2|\mathcal{E}(\mathcal{G})| - \widehat{Adv}_{\mathcal{G}}(v)$.
	
	\begin{IEEEproof}
		After step 1 in Algorithm \ref{alg:ST3}, which runs Algorithm \ref{alg:ST2} on the given information-flow graph $\mathcal{G}$, we get a tree $\mathscr{T}$ with $|S_{\mathscr{T}}| = adv_{\mathcal{G}}(v)$, (Proof in Lemma \ref{lem:T}), where $v$ is the chosen vertex with the maximum value of the modified advantage. For each $C_j, \ j \in [t]$ such that $Q(C_j)>0$, the tree $\mathscr{T}$ is updated as $\left(\mathscr{T} \setminus \mathscr{T}_{C_j}\right) \cup \{(v,u_j)\} \cup \mathscr{T}_j$, where $u_j$ is the bridge vertex in $C_j$. This operation removes all vertices in $\mathcal{V}(C_j)$ from $\mathscr{T}$ and joins the sub-tree $\mathscr{T}_j$ rooted at vertex $u_j$ by adding an edge $(v,u_j)$. 
		
		The sub-tree $\mathscr{T}_j$ is obtained by running \Call{UpdateTree}{} function on the star graph $\mathscr{S}_{u_j}^*(C_j)$. If  $\mathscr{T}_j$ is the same as $\mathscr{S}_{u_j}^*(C_j)$, the operation $\left(\mathscr{T} \setminus \mathscr{T}_{C_j}\right) \cup \{(v,u_j)\} \cup \mathscr{T}_j$ will increase $|S_{\mathscr{T}}|$ by $conn_{\mathcal{G}}(u_j,C_j)-conn_{\mathcal{G}}(v, \left(C_j \setminus \{u_j\}\right)$. Since, we remove all the vertices in $C_j$ from $\mathscr{T}$, all the advantage corresponding to vertices in $\left(\mathcal{P}_{\mathcal{G}}(v) \cup  \mathcal{O}_{\mathcal{G}}(v)\right)\cap \mathcal{V}(C_j)$ are lost. But after adding the sub-tree $\mathscr{T}_j$ which is returned by  \Call{UpdateTree}{$\mathscr{S}_{u_j}^*(C_j), \mathcal{G},  \mathcal{V}(C_j) \setminus (N_{\mathcal{G}}(v) \cup \{u_j\}), v$}, we will get advantage corresponding to $\left(\mathcal{P}_{\mathcal{G}}(u_j) \cup  \mathcal{O}_{\mathcal{G}}(u_j)\right)$. Since, $\mathcal{O}_{\mathcal{G}}(v) \cap \mathcal{O}_{\mathcal{G}}(u_j) = \mathcal{O}_{\mathcal{G}}(u_j)$, increment in $|S_{\mathscr{T}}|$ due to this step is $p_{\mathcal{G}}(u_j) - |\mathcal{P}_{\mathcal{G}}(v) \cap \mathcal{V}(C_j)|$. 
		
		Thus, for each component $C_j$ with $Q(C_j)>0$, $j \in [t]$, $|S_{\mathscr{T}}|$ is incremented by $conn_{\mathcal{G}}(u_j,C_j) - conn_{\mathcal{G}}(v, \left(C_j\setminus \{u_j\}\right) +  p_{\mathcal{G}}(u_j) - |\mathcal{P}_{\mathcal{G}}(v) \cap \mathcal{V}(C_j)|$ which is equal to $Q(C_j)$. Hence, $|S_{\mathscr{T}}| = adv_{\mathcal{G}}(v) + \sum_{j \in [t]}Q(C_j)\mathbbm{1}_{Q(C_j)>0}$ which is equal to $\widehat{Adv}_{\mathcal{G}}(v)$.
		
	\end{IEEEproof}
\end{lemma}
\begin{corollary}
	For the class of information-flow graphs satisfying the condition in \eqref{condA}, the index code based on the tree returned by Algorithm \ref{alg:ST3} has a reduced value of the total number of transmissions used than that based on the tree returned by Algorithm \ref{alg:ST1}. 
\end{corollary}
\begin{lemma}
	\label{lem:lmax}
	The index code based on the spanning tree returned by Algorithm \ref{alg:ST3} has $l_{max} \leq 2$.
	\begin{IEEEproof}
		The only change in Algorithm \ref{alg:ST2} from Algorithm \ref{alg:ST1} is that the criterion to choose the root vertex has been changed from maximizing $adv_{\mathcal{G}}$ to maximizing $\widehat{Adv}_{\mathcal{G}}$. The operation performed on the star graph to get the spanning tree hasn't changed because of which the proof that $l_{max} \leq 2$ for the  code based on the spanning tree obtained from Algorithm \ref{alg:ST1} continues to hold for the tree returned by Algorithm \ref{alg:ST2}. 
		
		In Algorithm \ref{alg:ST3}, a sub-tree is formed only for those components $C_j$ which have only one bridge vertex $u_j$ in it, which implies that only this vertex $u_j$ will have arcs to vertices in other components in $\mathcal{G}$. All the bridge vertices in $\mathcal{G}$ are connected to the root vertex $v$ and hence, are in level $L_1$ in the tree $\mathscr{T}$ returned by Algorithm A. Also, while calling the \Call{UpdateTree}{} function on $\mathscr{S}_{u_j}^*(C_j)$ for each $C_j$, all the vertices in the neighborhood of vertex $v$ which are in $C_j \setminus \{u_j\}$ are excluded from getting updated, hence fixing them in level $L_2$ in the final tree $\mathscr{T}$. 
	\end{IEEEproof}
\end{lemma}

Now, we introduce a class of information-flow graphs for which Algorithm \ref{alg:ST3} generates optimal spanning trees.
Consider an information-flow graph $\mathcal{G}$ for which the tree returned by Algorithm \ref{alg:ST3} is rooted at the vertex $v$ and each of the components in $\mathcal{G}^{'}(v) \setminus \mathcal{B}(v)$ contain only one bridge vertex in it. Let the set of these components be $\mathscr{C}(v) = \{C_1,C_2,\cdots,C_t\}$ with $\{u_j\} = \mathcal{V}(C_j) \cap \mathcal{V}_B(v)$, $j \in [t]$. Since each component has only one bridge vertex, $\mathcal{V}_B(v) = \{u_j\}_{j \in [t]}$. For such an information-flow graph $\mathcal{G}$, consider the following disjoint subsets of vertex-pairs, 

\begin{enumerate}
	\item $\mathcal{VP}_1 = \{(v,u_j), u_j \in \mathcal{V}_B(v)\}$,
	\item $\mathcal{VP}_2 =\{(u_j, w) : w \in \mathcal{V}(C_j) \setminus (\mathcal{P}_{\mathcal{G}}(u_j) \cup  \mathcal{O}_{\mathcal{G}}(u_j) \cup \{u_j\})$, for $j \in [t] \text{ s.t } Q(C_j) > 0\}$, and
	\item $\mathcal{VP}_3 = \{(x,y) : x \in (\mathcal{P}_{\mathcal{G}}(u_j) \cup  \mathcal{O}_{\mathcal{G}}(u_j)), \{y\} = N_{\mathcal{G}}(x) \setminus \{u_j\}$, for $j \in [t] \text{ s.t } Q(C_j) > 0\}$. 
\end{enumerate}

\begin{theorem}
	\label{Thm:Opt}
	Consider an information-flow graph $\mathcal{G}$ for which the tree $\mathscr{T}$ returned by Algorithm \ref{alg:ST3} is rooted at the vertex $v$ and all the $t \geq 2$ components $\{C_1,C_2,\cdots,C_t\}$, in $\mathcal{G}^{'}(v) \setminus \mathcal{B}(v)$ are such that  $\mathcal{V}(C_j) \cap \mathcal{V}_B(v) = \{u_j\}$ and $Q(C_j) >0$, $ \forall j \in [t]$.  If $\mathcal{G}$ satisfies either of the following two conditions, then $\mathscr{T}$ is optimal. 
	\begin{itemize}
		\item Condition 1: If a vertex pair $(i,j)$ does not belong to $\mathcal{VP}_1 \cup \mathcal{VP}_2 \cup \mathcal{VP}_3$, then $conn_{\mathcal{G}}(i,j) \leq 1$.
		\item Condition 2: If a vertex pair $(i,j)$ belongs to  $\mathcal{VP}_1 \cup \mathcal{VP}_2 \cup \mathcal{VP}_3$, then $conn_{\mathcal{G}}(i,j) = 2$.
	\end{itemize}
	\begin{IEEEproof}
		Since we require that a component should at least have two vertices, a vertex $u_j \in \mathcal{V}_B(v)$ cannot be present in either $\mathcal{P}_{\mathcal{G}}(v)$ or $\mathcal{O}_{\mathcal{G}}(v)$. Hence, in the tree $\mathscr{T}$ returned by Algorithm \ref{alg:ST3}, an edge of the form $(v,u_j)$ is always present. Further, since, for all $j \in [t]$, $Q(C_j) > 0$, a sub-tree $\mathscr{T}_j$ is formed for $C_j$, rooted at $u_j$, for every $C_j$. This sub-tree $\mathscr{T}_j$ will consist of edges of the form $(u_j,w)$, for all $w \in \mathcal{V}(C_j) \setminus \left(\mathcal{P}_{\mathcal{G}}(u_j) \cup  \mathcal{O}_{\mathcal{G}}(u_j) \cup \{u_j\}\right)$. This is because while running Algorithm \ref{alg:ST2} to generate $\mathscr{T}_j$, the only edges that will be removed from the star graph $\mathscr{S}^*_{u_j}(C_j)$ are of the form $(u_j,w)$ for $w \in \left(\mathcal{P}_{\mathcal{G}}(u_j) \cup \mathcal{O}_{\mathcal{G}}(u_j)\right)$. Also, for every $x \in \left(\mathcal{P}_{\mathcal{G}}(u_j) \cup  \mathcal{O}_{\mathcal{G}}(u_j)\right)$, Algorithm \ref{alg:ST2} adds an edge of the form $(x,y)$, where $\{y\} = N_{\mathcal{G}}(x) \setminus \{u_j\}$ to the sub-tree $\mathscr{T}_j$. 
		
		Thus, for a pair of vertices $(i,j)$, there exists an edge $(i,j)$ in $\mathscr{T}$ returned by Algorithm \ref{alg:ST3} if and only if $(i,j)$ belongs to one of $\mathcal{VP}_1$, $\mathcal{VP}_2$ and $\mathcal{VP}_3$. We will now prove that whenever $\mathcal{G}$ satisfies either one of the two conditions in Theorem, the spanning tree $\mathscr{T}$ returned by  Algorithm \ref{alg:ST3} is optimal. 
		
		Case 1: $\mathcal{G}$ satisfies Condition 1 - Let the number of vertices in $\mathcal{G}$ be $n$. The condition  that for all $(i,j)  \notin \mathcal{VP}_1 \cup \mathcal{VP}_2 \cup \mathcal{VP}_3$,  $conn_{\mathcal{G}}(i,j) \leq 1$ is equivalent to the condition that a double-arc can exist between two vertices $i$ and $j$ in $\mathcal{G}$ if and only if $(i,j)  \in \mathcal{VP}_1 \cup \mathcal{VP}_2 \cup \mathcal{VP}_3$. This implies that all double-arcs in $\mathcal{G}$ have a corresponding edge in $\mathscr{T}$ (and hence, the corresponding demands take one transmission each to decode), which in turn implies that the number of double-arcs in $\mathcal{G}$, denoted by $d({\mathcal{G}})$, is at most $n-1$.  The number of demands corresponding to the edges in $\mathscr{T}$ is $|S_{\mathscr{T}}| = (n-1)+d({\mathcal{G}})$. Therefore, the total number of transmissions taken by the index code obtained from $\mathscr{T}$ is given by $T =  2*|\mathcal{E}(\mathcal{G})| - (d({\mathcal{G}})+n-1)) = |\mathcal{E}(\mathcal{G})|  + |\mathcal{E}(\mathcal{G}_U)| - (n-1))$. This value of $T$ is equal to that  of the lower bound in Theorem \ref{Thm:LB1}, and hence, the spanning tree $\mathscr{T}$ is optimal. 
		
		Case 2: $\mathcal{G}$ satisfies Condition 2 - The condition that if $(i,j)  \in \mathcal{VP}_1 \cup \mathcal{VP}_2 \cup \mathcal{VP}_3$,  $conn_{\mathcal{G}}(i,j) =2$ implies that each edge in the spanning tree $\mathscr{T}$ has a corresponding double-arc in $\mathcal{G}$ which in turn implies that $d({\mathcal{G}}) \geq n-1$. In this case, $|S_{\mathscr{T}}| = 2(n-1)$ and hence, $T = 2*(|\mathcal{E}(\mathcal{G})| - 2(n-1)) + 2(n-1) = 2(|\mathcal{E}(\mathcal{G})| - (n-1))$ which is equal to the value of the lower bound in Theorem \ref{Thm:LB2}. Thus, the tree $\mathscr{T}$ is optimal. 
	\end{IEEEproof}
\end{theorem}

\section{Generalization to Union of Connected Components}
\label{sec:Gen}

\begin{figure}	
	\centering
	\scalebox{0.65}{\includegraphics{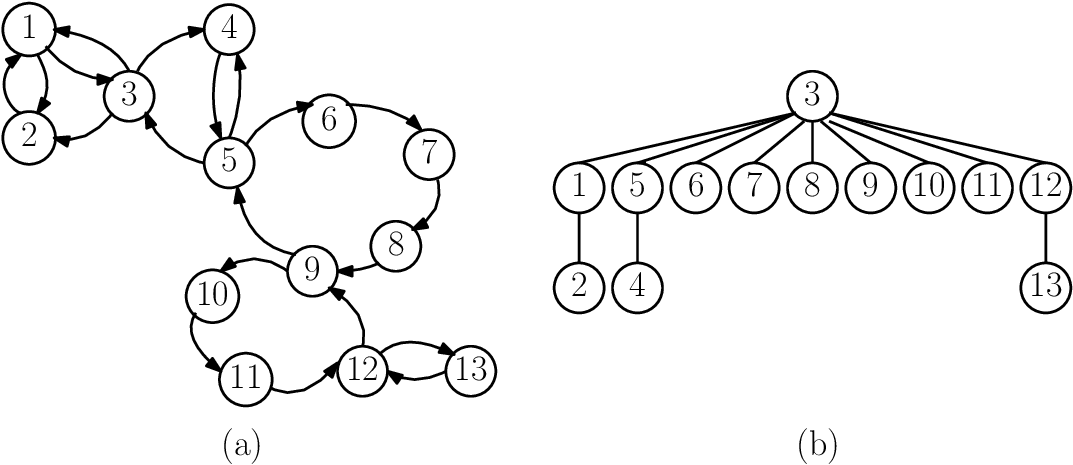}}
	\caption{ Information-flow graph $\mathcal{G}$  and the tree obtained from Algorithm \ref{alg:ST1} for it.}
	\label{fig:Ex_alg2}
	
	\centering
	\scalebox{0.65}{\includegraphics{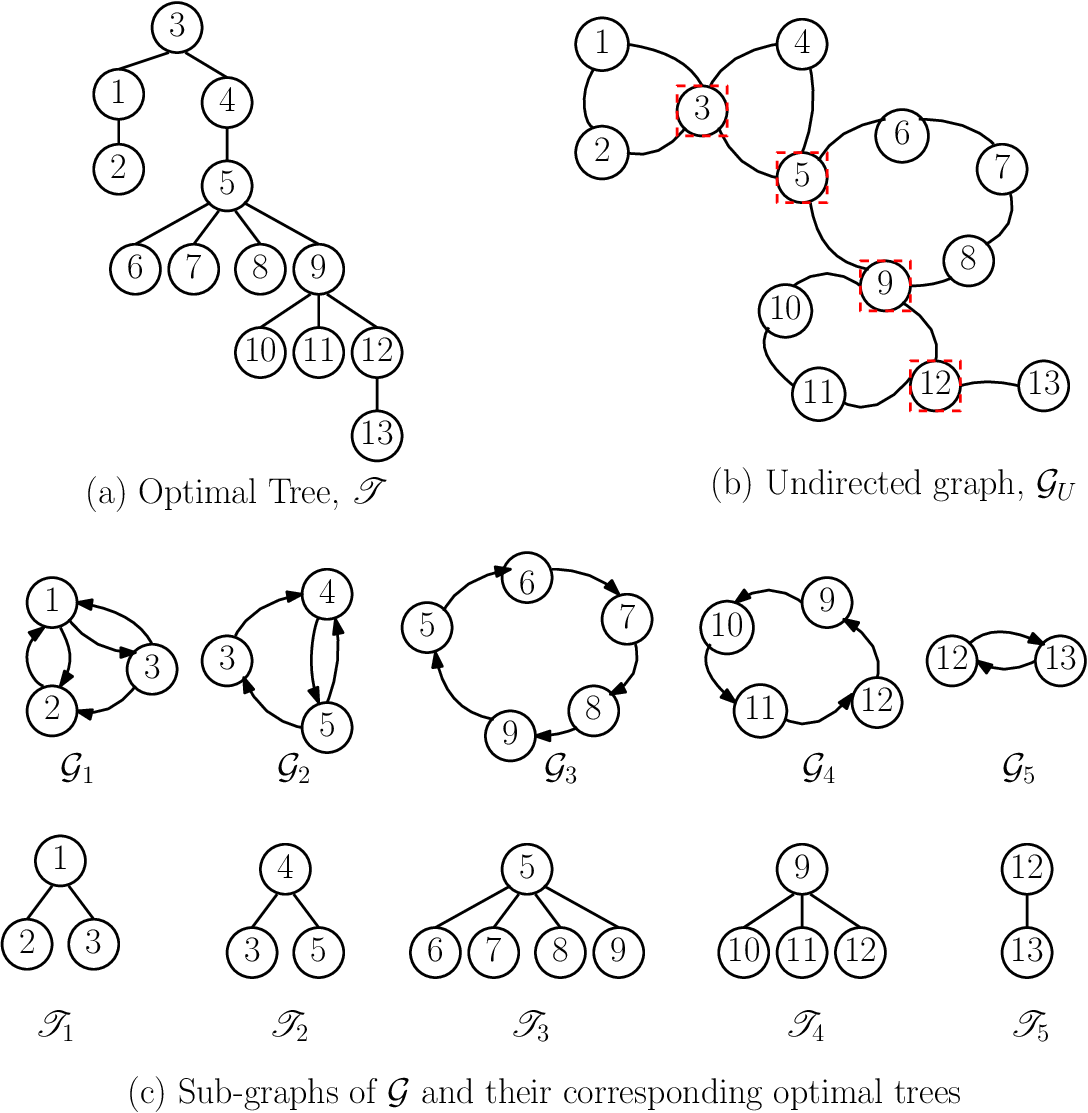}}
	\caption{Optimal spanning tree, undirected graph, sub-graphs and their optimal trees for the information-flow graph $\mathcal{G}$ in Figure \ref{fig:Ex_alg2}(a).}
	\label{fig:Ex_alg2_comp}
\end{figure}

Consider the information-flow graph $\mathcal{G}$ in Figure \ref{fig:Ex_alg2}(a). Running Algorithm \ref{alg:ST3} for this graph generates the spanning tree in Figure \ref{fig:Ex_alg2}(b), the index code based on which takes a total of $31$ transmissions for decoding the requested messages at the receivers. Consider the tree in Figure \ref{fig:Ex_alg2_comp}(a). For the index code based on this tree, the total number of transmissions required for decoding is $T = 27$. It can be verified that the tree $\mathscr{T}$ in Figure \ref{fig:Ex_alg2_comp}(a)  is optimal, w.r.t. the total number of transmissions used in decoding, for the information-flow graph $\mathcal{G}$ in Figure \ref{fig:Ex_alg2}(a) while satisfying the criterion that $l_{max} =2$. In the rest of this section, we explain how the tree in Figure \ref{fig:Ex_alg2_comp}(a) is obtained for this example and then generalize this procedure for information-flow graphs satisfying certain conditions. 

Consider the simplified undirected graph $\mathcal{G}_U$ shown in Figure \ref{fig:Ex_alg2_comp}(b)  corresponding to the information-flow graph $\mathcal{G}$ in Figure \ref{fig:Ex_alg2}(a). In the graph $\mathcal{G}_U$,  the vertices shown within dashed squares, namely $3, 5,9$, and $12$, are cut vertices or articulation points. Now, remove these cut vertices from $\mathcal{G}$ to get separate components, generate corresponding sub-graphs by duplicating the cut vertices and retaining the arcs to it in each component, and find the spanning tree for each of these components by running Algorithm \ref{alg:ST3}. For the information-flow graph $\mathcal{G}$, the set of sub-graphs $ \{\mathcal{G}_1,\mathcal{G}_2,\mathcal{G}_3,\mathcal{G}_4,\mathcal{G}_5\}$ and their corresponding trees $\{\mathscr{T}_1, \mathscr{T}_2,  \mathscr{T}_3, \mathscr{T}_4, \mathscr{T}_5\}$ are shown in Figure \ref{fig:Ex_alg2_comp}(c). Finally, join these trees by overlapping at the cut vertices to obtain the tree for the graph $\mathcal{G}$. 

As explained in Lemma \ref{lem:T}, the condition $\exists \ j \in \mathcal{V} \setminus \{i\}$ s.t. $N_{\mathcal{G}}(j) \setminus \{i\} = \{k\}$ and $conn_{\mathcal{G}}(j,k) > conn_{\mathcal{G}}(i,j)$ checked in \Call{UpdateTree}{} for modifying the star graph $\mathscr{S}^*_{i}(\mathcal{G})$ can be split into two equivalent conditions, the first of which is given as $j \notin N_{\mathcal{G}}(i)$ and $N_{\mathcal{G}}(j) = \{k\}$. 
For each vertex $j$ satisfying this condition, its neighbor vertex $k$ acts as a cut vertex in $\mathcal{G}_U$, giving a component with just the vertices $k$ and $j$ and a double-arc between them for which the tree will be the edge $(k,j)$. Hence, the operation performed by the Algorithm \ref{alg:ST2} on such vertices is no longer needed. Therefore, we modify the while condition in the function \Call{UpdateTree}{} as \textbf{while} $\exists \ j \in \mathcal{V}^{'} \text{ s.t. } N_{\mathcal{G}}(j) = \{i,k\}$ and $conn_{\mathcal{G}}(j,k) > conn_{\mathcal{G}}(i,j)$. With this change, the function \Call{UpdateTree}{} is restated as \Call{UpdateTreeNew}{} and Algorithm \ref{alg:ST2} with a call to the modified function is given as Algorithm \ref{alg:ST4} below.

\begin{algorithm}
	\caption{Modified version of Algorithm \ref{alg:ST2}}
	
	\begin{algorithmic}[1]
		
		\Require Information-flow graph, $\mathcal{G} = (\mathcal{V},\mathcal{E})$
		\Ensure Tree $\mathscr{T}$
		\State Find the set of all vertices with maximum modified advantage, $\mathcal{A} \coloneqq \arg\max\limits_{v \in \mathcal{V}}(\widehat{Adv}_{\mathcal{G}}(v)) $. 
		\State Find the vertices in $\mathcal{A}$ with maximum degree,  $\mathcal{A}_{\Delta} \coloneqq \arg\max\limits_{v \in \mathcal{A}}(\deg_{\mathcal{G}}(v))$. 
		\State Pick an $i$ from $\mathcal{A}_{\Delta}$.
		\State $\mathscr{T} \leftarrow \mathscr{S}^*_{i}(\mathcal{G})$. 
		\If{$\Delta(\mathcal{G}) < \widehat{Adv}_{\mathcal{G}}(i) $}
		
		\State $\mathcal{V}^{'}= \mathcal{V} \setminus \{i\}$.
		\State \Call{UpdateTreeNew}{$\mathscr{T}$, $\mathcal{G}$, $\mathcal{V}^{'}$, $i$}
		\EndIf	
	\end{algorithmic}
	\label{alg:ST4}
\end{algorithm}

\begin{algorithm}
	\caption*{Modified Version of the Function \Call{UdpateTree}{$\mathscr{T}$, $\mathcal{G}$, $\mathcal{V}^{'}$, $i$}}
	\begin{algorithmic}[1]
		\Procedure{UpdateTreeNew}{$\mathscr{T}$, $\mathcal{G}$, $\mathcal{V}^{'}$, $i$}
		\While {  $\exists \ j \in \mathcal{V}^{'} $ s.t. $N_{\mathcal{G}}(j) = \{i,k\}$ and $conn_{\mathcal{G}}(j,k) > conn_{\mathcal{G}}(i,j)$} 
		\State $\mathcal{E}(\mathscr{T}) = \big\{\mathcal{E}(\mathscr{T}) \setminus \{(i,j)\} \big\}  \cup \{(k,j)\} $.
		\State $\mathcal{V}^{'} = \mathcal{V}^{'} \setminus \{j,k\}$.
		\EndWhile
		\State \textbf{return} $\mathscr{T}$
		\EndProcedure
	\end{algorithmic}
	\label{func:mod_update}	
\end{algorithm}

Now we propose the following Algorithm \ref{alg:ST_Gen} for generating an improved spanning tree for information-flow graphs with articulation points. Note that, for a given information-flow graph $\mathcal{G}$, we are only considering articulation points and not strong articulation points (SAPs) in $\mathcal{G}$ as, in general, removal of SAPs will not disconnect the graph but will only increase the number of strongly connected components. We are looking for cut vertices or articulation points in the graph, which, when removed, will disconnect the graph. Let $S$ denote the set of articulation points, and upon removal of the vertices in $S$, let there be $k$ components, $C_1,C_2,\cdots,C_t$ in $\mathcal{G} \setminus \{S\}$. Corresponding to each component $C_i$, $i \in [t]$, let $S_{C_i}$ denote the subset of vertices in $S$ which has a neighbor in $C_i$, i.e., $S_{C_i} \triangleq \{v \in S \text{ s.t. } N_{\mathcal{G}}(v) \cap \mathcal{V}(C_i) \neq \emptyset\}$. Algorithm \ref{alg:ST_Gen} considers the vertex-induced sub-graph  $\mathcal{G}_i \subseteq \mathcal{G}$ on the vertex set $\mathcal{V}(C_i) \cup S_{C_i}$, for each $i \in [t]$. 


\begin{algorithm}
	\caption{Generate a spanning tree for information-flow graphs with cut vertices }
	\label{alg:ST_Gen}
	\begin{algorithmic}[1]
		
		\Require Information-flow graph, $\mathcal{G} = (\mathcal{V},\mathcal{E})$
		\Ensure Tree $\mathscr{T}$
		
		\State Let $S = \{v \in \mathcal{V} \ | \ v \text{ is an articulation point in } \mathcal{G}_U \}$
		\State Find all the components of $\mathcal{G} \setminus S$, say, $C_1, C_2, \cdots, C_t$, and form the corresponding sub-graphs $\{\mathcal{G}_1,\mathcal{G}_2,\cdots,\mathcal{G}_t\}$. 
		
		\For{each $i \in [t]$}
		\State Run Algorithm \ref{alg:ST4} for the sub-graph $\mathcal{G}_i$ to get the tree $\mathscr{T}_i$. 
		\EndFor
		
		\State Combine all trees $\{\mathscr{T}_i\}_{i \in [t]}$ together by overlapping at the vertices in $S$ to get the tree $\mathscr{T}$. 
		
	\end{algorithmic}
\end{algorithm}

\begin{remark}
	The total number of transmissions used in decoding the index code based on the tree $\mathscr{T}$ returned by Algorithm \ref{alg:ST_Gen} is equal to the sum of the transmissions used in decoding the codes based on the component trees $\mathscr{T}_1, \mathscr{T}_2, \cdots, \mathscr{T}_t$. 
\end{remark}
\begin{lemma}
	\label{lem:lmax_Gen}
	For an index code based on the tree $\mathscr{T}$ returned by Algorithm \ref{alg:ST_Gen}, the maximum number of transmissions used in decoding any requested message at any receiver, $l_{max}$, is $2$. 
	
	\begin{IEEEproof}
		The trees $\mathscr{T}_1, \mathscr{T}_2, \cdots, \mathscr{T}_t$, returned by Algorithm \ref{alg:ST4}  for the sub-graphs $\mathcal{G}_1, \mathcal{G}_2, \cdots, \mathcal{G}_t$ of the information-flow graph $\mathcal{G}$, satisfy the condition that $l_{max} = 2$ as seen from Lemma \ref{lem:lmax}. Hence, the receivers corresponding to the vertices in $\mathcal{V}(\mathcal{G}) \setminus S$ all take at most two transmissions to decode a requested message since these vertices do not have a neighbor outside of their respective sub-graphs. Now consider a receiver $R_i$ such that $i \in S$. Let the removal of the vertex $i$ from $\mathcal{G}$ result in $k$ components corresponding to which the sub-graphs are $\mathcal{G}_{i_1}, \mathcal{G}_{i_2}, \cdots, \mathcal{G}_{i_k}$. The neighbors of $i$ in each of these sub-graphs are at a maximum distance of $2$ from $i$ in their corresponding trees. Since the trees corresponding to these $k$ sub-graphs are all joined together by overlapping at the vertex $i$, the distance of the neighbors of $i$ in the final tree is the same as that in the component trees. Hence, for $R_i$, the maximum number of transmissions used in decoding any single message is at most two. Therefore, $l_{max}$ for the tree $\mathscr{T}$ returned by Algorithm \ref{alg:ST_Gen} is two. 
	\end{IEEEproof}
\end{lemma}

\begin{theorem}
	\label{Thm:C2_Opt_Gen}
	If all the trees $\{\mathscr{T}_i\}_{i \in [t]}$ corresponding to the sub-graphs of a given information-flow graph $\mathcal{G}$ in Algorithm \ref{alg:ST_Gen} are optimal, then the tree $\mathscr{T}$ returned by it is also optimal. 
	\begin{IEEEproof}		
		Let there be only one articulation point, say $v$, in $\mathcal{G}_U$, the removal of which results in two components and the sub-graph corresponding to which are denoted as $\mathcal{G}_1$ and $\mathcal{G}_2$. Assume to the contrary that there is a bandwidth-optimal index code $\mathcal{C}^{'}$ for $\mathcal{G}$, satisfying $l_{max}(\mathcal{C}) = 2$, which uses a total number of transmissions $T^{'}$ for decoding which is less than the total number of transmissions $T$ used in the code $\mathcal{C}$ based on the tree $\mathscr{T}$ returned by Algorithm \ref{alg:ST_Gen}.
		
		\textbf{Claim}: When the information-flow graph $\mathcal{G}$ on $n$ vertices is strongly connected, every transmission in any index code of length $n-1$, is of the form $x_i + x_j$, $i,j \in [n]$. 
		
		Assuming the claim to be true, the transmissions in $\mathcal{C}^{'}$ are also of the form $x_i +x_k$. Consider a transmission $x_{j_1} + x_{k_1}$ in $\mathcal{C}^{'}$, where $j_1 \in \mathcal{V}(\mathcal{G}_1) \setminus \{v\}$ and $k_1 \in \mathcal{V}(\mathcal{G}_2) \setminus \{v\}$. Any receiver cannot use such a transmission by itself to decode a requested message since no receiver demanding $x_{j_1}$ knows $x_{k_1}$ and vice-versa. Hence, it requires two other transmissions of the form $x_{j_1} + x_{j_2}$ and $x_{k_1} + x_{k_2}$, where $j_1, j_2 \in \mathcal{V}(\mathcal{G}_1) \setminus \{v\}$ and $k_1, k_2 \in \mathcal{V}(\mathcal{G}_2) \setminus \{v\}$, to be useful. However, these three transmissions can be replaced with just the two $x_{j_1} + x_{j_2}$ and $x_{k_1} + x_{k_2}$. Since $\mathcal{C}^{'}$ is bandwidth-optimal, it doesn't contain transmissions of the form $x_{j_1} + x_{k_1}$ in $\mathcal{C}^{'}$, where $j_1 \in \mathcal{V}(\mathcal{G}_1) \setminus \{v\}$ and $k_1 \in \mathcal{V}(\mathcal{G}_2) \setminus \{v\}$.  Therefore, the allowed transmissions in $\mathcal{C}^{'}$ are of the form $x_{j_1} + x_{j_2}$, $x_{j_1} + x_v$, $x_v + x_{k_1}$ and $x_{k_1} + x_{k_2}$, where, $j_1, j_2 \in \mathcal{V}(\mathcal{G}_1) \setminus \{v\}$ and $k_1, k_2 \in \mathcal{V}(\mathcal{G}_2) \setminus \{v\}$. Hence, the transmissions in $\mathcal{C}^{'}$ can be partitioned into two codes, $C_1^{'}$ containing transmissions of the form $x_{j_1}+ x_{j_2}$ and $x_{j_1} + x_v$ and $C_2^{'}$ containing transmissions of the form $x_{k_1}+ x_{k_2}$ and $x_{k_1} + x_v$. It can be seen that $C_1^{'}$ is an index code for the sub-graph $\mathcal{G}_1$ whereas $C_2^{'}$ is an index code for the sub-graph $\mathcal{G}_2$. Let the transmissions used in decoding for $C_1^{'}$ and $C_2^{'}$ be $T_1^{'}$ and $T_2^{'}$, respectively which implies that $T^{'} = T_1^{'} + T_2^{'}$. 
		
		Let the codes based on the trees $\mathscr{T}_1$ and $\mathscr{T}_2$ obtained by running Algorithm \ref{alg:ST4} on the sub-graphs $\mathcal{G}_1$ and $\mathcal{G}_2$ be denoted as $C_1$ and $C_2$. The total number of transmissions used in decoding $C_1$ and $C_2$ are denoted $T_1$ and $T_2$, respectively, which gives $T = T_1 + T_2$. By assumption  $T^{'}$ is less than $T$, i.e., $T_1^{'} + T_2^{'} < T_1 + T_2$ which implies that $T_1^{'} < T_1$ or $T_2^{'} < T_2$ or both. However, none of this is possible since we assumed that the component trees $\mathscr{T}_1$ and $\mathscr{T}_2$ are optimal w.r.t. the total number of transmissions used in decoding. Hence, there cannot exist such a $\mathcal{C}^{'}$.
	\end{IEEEproof}
\end{theorem}

\textbf{Proof of Claim}: First, we prove that no transmission contains just a single message in an index code of length $n-1$. Suppose a message, say $x_1$, is transmitted alone without coding. Then, to satisfy the message requests of $R_1$, there exists at least one transmission either of the form $x_1 + x_j$ or $x_j$, for some $j \in \mathcal{V}(\mathcal{G}) \setminus \{1\}$. Suppose a transmission $x_j$ exists, where $x_j$ is demanded by $R_1$. Since $\mathcal{G}$ is strongly connected, there exists another message demanded by either  $R_1$ or $R_j$ to satisfy which, again, there should be a transmission of the form $x_k$ or $x_1 + x_k$ or $x_j + x_k$. Consider the other case, where after transmitting $x_1$ alone, we transmit a message of the form $x_1 + x_j$. Since $\mathcal{G}$ is strongly connected, using the same argument as in the case where $x_j$ is transmitted alone, there exists a transmission of the form $x_k$ or $x_1 + x_k$ or $x_j + x_k$. This argument continues, and a new message is involved at every step. It can be verified that at $(n-1)^{\text{th}}$ step, it will still be required to transmit another message due to the connectedness of the graph, thus leading to a total of $n$ transmissions in the index code which is not optimal. Since in a single uniprior ICP, each receiver knows only one message \textit{a priori}; it can be seen that there cannot exist a bandwidth-optimal index code that has a transmission that is a linear combination of three or more messages while still satisfying $l_{max} = 2$.

\begin{corollary}
	If all the sub-graphs of a given information-flow graph $\mathcal{G}$ in Algorithm \ref{alg:ST_Gen} satisfy the conditions in Theorem \ref{Thm:Opt1} or Theorem \ref{Thm:Opt2} or Theorem \ref{Thm:Opt}, then the tree $\mathscr{T}$ returned by Algorithm \ref{alg:ST_Gen} is optimal.
\end{corollary}


\section{Conclusion}
\label{sec:Conc}

\begin{table}
	\centering	
\begin{tabular}{|c|p{5cm}|c|p{6cm}|}
		\hline
		Algorithm & Description & Subroutines Invoked & Total number of transmissions used, $T$ \\
		\hline
		Algorithm \ref{alg:ST1}&  Generate a spanning tree that improves upon the optimal star graph & \Call{UpdateTree}{} & $T = 2|\mathcal{E}(\mathcal{G})| - \max\limits_{v \in \mathcal{V}(\mathcal{G})}\left(adv_{\mathcal{G}}(v)\right)$ \\
		\hline
		Algorithm \ref{alg:ST2}&  Modified version of Algorithm \ref{alg:ST1}  & \Call{UpdateTree}{} & $T = 2|\mathcal{E}(\mathcal{G})| - \max\limits_{v \in \mathcal{V}(\mathcal{G})}\left(\widehat{Adv}_{\mathcal{G}}(v)\right)$ \\
		\hline
		Algorithm \ref{alg:ST3}&  Generate a spanning tree for IFGs with bridges  & Algorithm \ref{alg:ST2} & $T = 2|\mathcal{E}(\mathcal{G})| - \max\limits_{v \in \mathcal{V}(\mathcal{G})}\left(adv_{\mathcal{G}}(v)\right)$ \\
		\hline
		Algorithm \ref{alg:ST4}&  Modified version of Algorithm \ref{alg:ST2} &\Call{UpdateTreeNew}{} &  $T = 2|\mathcal{E}(\mathcal{G})| - \max\limits_{v \in \mathcal{V}(\mathcal{G})}\left(adv_{\mathcal{G}}(v)\right)$ \\
		\hline
		Algorithm \ref{alg:ST_Gen}&  Generalization of Algorithm \ref{alg:ST4} for IFGs with cut vertices& Algorithm \ref{alg:ST4} & $T = 2|\mathcal{E}(\mathcal{G})| -\sum\limits_{i = 1}^{t}\left(\max\limits_{v \in \mathcal{V}(\mathcal{G}_i)}\left(adv_{\mathcal{G}_i}(v)\right)\right)$, where $\mathcal{G}_1,\cdots,\mathcal{G}_t$ are the $t$ sub-graphs. \\
		\hline
	\end{tabular}
	\caption{Summary of Proposed Algorithms.}
	\label{Tab:Sumamry}
\end{table}

Inspired by the min-max probability of error criterion in \cite{TRAR}, which gave a rule for the selection of an index code from the class of bandwidth-optimal linear index codes for binary-modulated index code transmission over Rayleigh fading broadcast channels, we considered single uniprior index coding problems over continuous-output with binary-modulated transmissions. For a further selection from the class of bandwidth-optimal index codes taking at most two transmissions to decode a requested message, we introduced minimizing the average probability of error as a criterion, where the average is taken over all message requests by all the receivers. 

For a given single uniprior index coding problem, we only optimize the average probability of error from the set of index codes satisfying the following two criteria. 1) The index code is bandwidth-optimal, and 2)  the index code satisfies the min-max probability of error criterion in \mbox{\cite{TRAR}}.  This is a reasonable approach for the class of  single uniprior index coding problems considered in this paper because of the following reasons. 
\begin{itemize}
	\item The minimum length, as well as the construction of a minimum-length index code for each strongly connected component, are deterministic, and there are no computational costs involved in finding them.  
	\item The index codes satisfying the min-max probability of error are shown to be the star graphs.
	\item For a strongly connected component of the information flow graph on $n$ vertices, the minimum length of the index code is $n-1$ which implies that with $n-2$ transmissions or less it is not possible to satisfy the demands of all the receivers. With the $n$ transmissions, the code is trivial since all messages can be transmitted independently without code. Hence, the only case worth discussing is the class of bandwidth-optimal codes.
\end{itemize}

It was shown that minimizing the average probability of error is equivalent to minimizing the total number of transmissions used in decoding the message requests at all the receivers. Algorithms for generating spanning trees for strongly connected information-flow graphs, representing single uniprior index coding problems, which resulted in index codes that minimized the total number of transmissions used in decoding, were given. A summary of all the algorithms in this paper is provided in Table \mbox{\ref{Tab:Sumamry}}. Two lower bounds were given for the minimum possible value of the total number of transmissions used for a given set of parameters of the ICP. A few classes of single uniprior ICPs for which these lower bounds are tight were also identified. 

It has been manually verified that for all strongly connected information-flow graphs on five or fewer vertices, Algorithm \ref{alg:ST1} generates an optimal spanning tree. However, other than for the two classes of graphs in section \ref{sec:LBOpt}, we have not exactly characterized the set of information-flow graphs for which Algorithm \ref{alg:ST1} generates optimal spanning trees. Similarly, for Algorithm \ref{alg:ST3}, only one class of graphs has been identified for which it generates optimal spanning trees. Hence, an interesting topic for research will be characterizing other classes of graphs for which the algorithms in this paper give optimal spanning trees. Another problem of interest is improving the algorithms in this paper so that optimal spanning trees can be generated for all strongly connected graphs.
	
	\section{Appendix}
	\textbf{Graph Theoretic Preliminaries} : 
	The following is a list of some basic graph theoretic definitions \cite{RD, DBW} used in this paper.
A graph $G$ is a triple consisting of a vertex set $V(G)$, an edge set $E(G)$, and a relation that associates with each edge two vertices called its endpoints. A \emph{directed} graph is a graph with a direction associated with each edge in it. This implies that an edge $(u,v)$  directed from the vertex $u$ to the vertex $v$ and an edge $(v,u)$ from $v$ to $u$ are two different edges in a directed graph, whereas in an undirected graph, $(u,v)$ and $(v,u)$ mean the same edge between the endpoints $u$ and $v$.	The edges in a directed graph are also called arcs. Two vertices $u$ and $v$ are \emph{adjacent} or \emph{neighbors} in an undirected graph $G$ if there exists an edge $(u,v)$ in $G$. In a directed graph $G$, the out-neighborhood of a vertex $u$, denoted by $N_G^+(u)$, is the set of vertices $\{v:(u,v) \in E(G)\}$ and its in-neighborhood, denoted as $N_G^-(u)$, is the set of vertices $\{v:(v,u) \in E(G)\}$. The set of neighbors of a vertex $v \in V(G)$ is denoted as $N_G(v)$ and for a directed graph $N_G(v) = N_G^+(u) \cup N_G^-(u)$. A graph $G'$ is called a sub-graph of the graph $G$, written as $G' \subseteq G$, if $V(G') \subseteq V(G)$ and $E(G') \subseteq E(G)$. If $G' \subseteq G$ and $G'$ contains all the edges $(x,y) \in E(G)$ with $x, y \in V(G')$, then G' is called an \emph{induced} sub graph of $G$. The \emph{degree} of a vertex $v$, $\deg_{G}(v)$, is the number of edges incident at it, which is equal to the number of neighbors of the vertex $v$ in the graph $G$. A vertex of degree $0$ is called an \emph{isolated} vertex. For a vertex $v$ in a directed graph $G$, its in-degree, denoted as $in-\deg_{G}(v)$ is the number of vertices in its in-neighborhood and its out-degree $out-\deg_{G}(v)$ is the number of vertices in its out-neighborhood and hence its degree, $\deg_{G}(v)$ is equal to the sum of its in-degree and out-degree in $G$. A \emph{path} is a non-empty graph $P = (V,E)$ of the form $V = \{x_0,x_1,\cdots,x_{k-1},x_k\}$ and $E = \{(x_0,x_1), (x_1,x_2),\cdots,(x_{k-1},x_k)\}$, where all the $x_i$s are all distinct. A \emph{cycle} is a path with the same first and last vertices being the same. An undirected graph $G$ is called \emph{connected} if it is non-empty and any two of its vertices are linked by a path in $G$. A maximal connected sub-graph of $G$ is a \emph{component} of $G$. A directed graph $G$ is said to be \emph{strongly connected} if there exists a path from $u$ to $v$ and another path from $v$ to $u$ for every pair of vertices $u, v \in V(G)$. A \emph{strongly connected component} of a directed graph $G$ is a sub-graph of $G$ that is strongly connected and is maximal with this property. An undirected graph is said to be a \emph{tree} if it is connected and does not have any cycles. A tree on $n$ vertices has $n-1$ edges. A \emph{rooted tree} is a tree in which one vertex, called the \emph{root vertex}, is distinguishable from the others. A rooted tree with root vertex $v$ is said to be a \emph{star graph} with vertex $v$ as \emph{head} if every edge in the tree has $v$ as one of its endpoints. For an undirected graph $G$, a \emph{spanning tree} is a sub-graph which is a tree and includes all the vertices in $G$. For a pair of vertices $u,v \in V(G)$ in a connected graph $G$, the \emph{distance} between $u$ and $v$, denoted $dist(u,v)$, is the length of the shortest path between $u$ and $v$. The \emph{diameter} of a graph is the length of the shortest path between a pair of vertices that are at maximum distance from each other in $G$, i.e., $diam(G) = \max\limits_{u \in V(G)}\{\max\limits_{v \in V(G)}(dist(u,v))\}$. A \emph{complete graph} on $n$ vertices, denoted $K_n$, is the undirected graph on $n$ vertices, with every pair of vertices being connected by an edge. An edge in an undirected graph is said to be a \emph{bridge} if the removal of it disconnects the graph. A vertex $v \in V(G)$ is said to be a \emph{cut vertex} or an \emph{articulation point} in a connected graph $G$ if the removal of the vertex $v$ and the edges incident on it from $G$ disconnects $G$. A vertex $v \in V(G)$ of a strongly connected graph $G$ is said to be a \emph{strong articulation point} (SAP) of $G$ if the removal of it increases the number of strongly connected components in $G$. 
%
%
	\section*{Acknowledgment}
	This work was supported partly by the Science and Engineering Research Board (SERB) of Department of Science and Technology (DST), Government of India, through J.C. Bose National Fellowship to Prof. B. Sundar Rajan and by the C V Raman postdoctoral fellowship awarded to Dr. Charul Rajput. 
	
	

\end{document}